\documentclass[12pt,preprint]{aastex}
\newcommand{\Msun}{\ifmmode\mbox{M}_{\odot}\else$\mbox{M}_{\odot}$\fi}
\newcommand{\Rsun}{\ifmmode\mbox{R}_{\odot}\else$\mbox{R}_{\odot}$\fi}
\newcommand{\Mearth}{\ifmmode\mbox{M}_{\oplus}\else$\mbox{M}_{\oplus}$\fi}
\newcommand{\Rearth}{\ifmmode\mbox{R}_{\oplus}\else$\mbox{R}_{\oplus}$\fi}

\newcommand{\seven}{RXS~J170849.0$-$400910}
\newcommand{\eight}{1E~1841$-$045}

\shorttitle{Glitches in AXPs}
\shortauthors{Dib et al.}
\begin{document}
\title{Glitches in Anomalous X-ray Pulsars}
\author{Rim~Dib\altaffilmark{1},
        Victoria~M.~Kaspi\altaffilmark{1}, and
	Fotis~P.~Gavriil\altaffilmark{2}}
\altaffiltext{1}{Department of Physics, McGill University,
                 Montreal, QC H3A~2T8}
\altaffiltext{2}{NASA Goddard Space Flight Center, Greenbelt, MD.}
\begin{abstract}
We report on 8.7 and 7.6~yr of {\it Rossi X-ray Timing Explorer (RXTE)}
observations of the Anomalous X-ray Pulsars (AXPs) \seven\ ~and \eight,
respectively.  These observations, part of a larger {\it RXTE} AXP
monitoring program, have allowed us to study the long-term timing, pulsed
flux, and pulse profile evolution of these objects.  We report on four new
glitches, one from \seven\ and three from \eight. One of the glitches from
\eight\ is among the largest ever seen in a neutron star in terms of
fractional frequency increase.  With nearly all known persistent AXPs now
seen to glitch, such behavior is clearly generic to this source class.  We
show that in terms of fractional frequency change, AXPs are among the most
actively glitching neutron stars, with glitch amplitudes in general larger
than in radio pulsars.  However, in terms of absolute glitch amplitude, AXP
glitches are unremarkable.  We show that the largest AXP glitches observed
thus far have recoveries that are unusual among those of radio pulsar
glitches, with the combination of recovery time scale and fraction yielding
changes in spin-down rates following the glitch similar to, or larger than,
the long-term average.  We also observed a large long-term fractional
increase in the magnitude of the spin-down rate of \eight, following its
largest glitch, with $\Delta\dot{\nu}/\dot{\nu}=0.1$. These observations are
challenging to interpret in standard glitch models, as is the frequent
occurence of large glitches given AXPs' high measured temperatures.  We
speculate that the stellar core may be involved in the largest AXP glitches.
Furthermore, we show that AXP glitches appear to fall in two classes:
radiatively loud and radiatively quiet.  The latter, of which the glitches
of \seven\ and \eight\ are examples, show little evidence for an
accompanying radiative event such as a sudden flux increase or pulse profile
change.  We also show, however, that pulse profile and pulsed flux changes
are common in these AXPs, but do not apprear closely correlated with any
timing behavior.
\end{abstract}
\section{Introduction}
The past decade has seen significant progress in our knowledge of the
observational properties of Anomalous X-ray Pulsars \citep[AXPs; see][for
recent reviews]{wt06,kas07}.  From a timing point of view, the presence of
binary companions has been practically ruled out \citep{mis98,wdf+99}, and
subsequently their potential for great rotational stability was demonstrated
\citep{kcs99}, thereby allowing the discovery that AXPs can exhibit spin-up
glitches \citep{klc00,kg03,dis+03}, and large (factor of $\sim$10) torque
variations \citep{gk04}.  From a radiative point of view, AXPs are now known
to show a variety of different variability phenomena, including long-lived
flares \citep{gk04}, short SGR-like bursts \citep{gkw02,gkw04,wkg+05}, large
outbursts (\citealt{kgw+03,ims+04,icd+07,dkgw07,tkgg06}; Tam et al.,
submitted), and slow, low-level flux and pulse profile variability
(\citealt{dkg07}; Gonzalez et al., submitted). Spectrally, though previously
studied only in the soft X-ray band, AXPs are now seen in the radio band
\citep{crh+06}, through the mid- \citep{wck06} and near-IR
\citep[e.g.,][]{ics+02,wc02,hvk04,tkvd04,rit+04,dvk05}, in the optical range
\citep[e.g.,][]{km02,dmh+05}, up to hard X-ray energies \citep{khdc06}.  The
evidence thus far argues strongly that AXPs, like their close cousins, the
Soft Gamma Repeaters, are magnetars -- young, isolated neutron stars powered
by a large magnetic energy reservoir, with surface fields of
$>10^{14}-10^{15}$~G \citep{td96a,tlk02}.

In spite of this progress, however, many aspects of AXPs remain mysterious.
Particularly so are their variability properties. What is the origin of the
variety of different types of variability? Although bursts can be explained
as sudden crustal yields, slower evolution \citep[e.g.,][]{gk04,dkg07} has
been suggested to be due to slow magnetospheric twists \citep{tlk02}.  Some
support for this picture has been argued to come from observed correlations
between flux and spectral hardness \citep{wkt+04,roz+05,cri+07}, although
\citet{og07} argue that such a correlation need not originate uniquely from
the magnetosphere and could be purely thermal.  At least some radiative
variability has been seen to be correlated with timing behavior. The best
example of this occured in the 2002 outburst of AXP 1E~2259+586 in which the
pulsar suffered a large spin-up glitch apparently simultaneously with a
major X-ray outburst \citep{kgw+03,wkt+04}.  \citet{icd+07} describe a
similar radiative outburst in AXP CXOU J164710.2$-$455216, and report a
large contemporaneous glitch, as did \citet{dkgw07} recently for AXP
1E~1048.1$-$5937.  By contrast, AXP \seven\ exhibited two glitches with no
evidence for a corresponding radiative event \citep{klc00,kg03} although
\citet{dis+03} suggested possible low-level pulse profile changes associated
with the second glitch. \cite{cri+07} also suggested that observed flux and
spectral changes may be associated with glitches and predicted a third
glitch would be observed after mid-2005 on the basis of an observed flux
increase and apparently correlated spectral changes.

Here we report on 8.7 and 7.6~yr of monitoring of \seven\ and \eight,
respectively, using the Proportional Counter Array (PCA) aboard the {\it
Rossi X-ray Timing Explorer} ({\em RXTE}).  We report the discovery of one
new glitch and three new glitch candidates in \seven\, as well as three new
glitches in \eight, including one of the largest glitches, in terms of
fractional frequency increase, thus far observed in any neutron star.  We
also present pulsed flux time series for \seven\ and \eight\ which reveal
little or no evidence for correlated changes with glitches, although \seven\
shows low-level pulsed flux variability at many epochs.  We also report a
pulse profile evolution analysis which shows that both pulsars' profiles are
evolving slowly with time, though in neither case does this evolution show a
clear correlation with timing behavior.  These results demonstrate that AXPs
\seven\ and \eight\ are frequent glitchers. They also demonstrate that
although AXP timing glitches can occur simultaneously with significant
long-lived radiative enhancements, they need not always do so.

\section{Observations}
\label{sec:observations}

\begin{figure}
\includegraphics{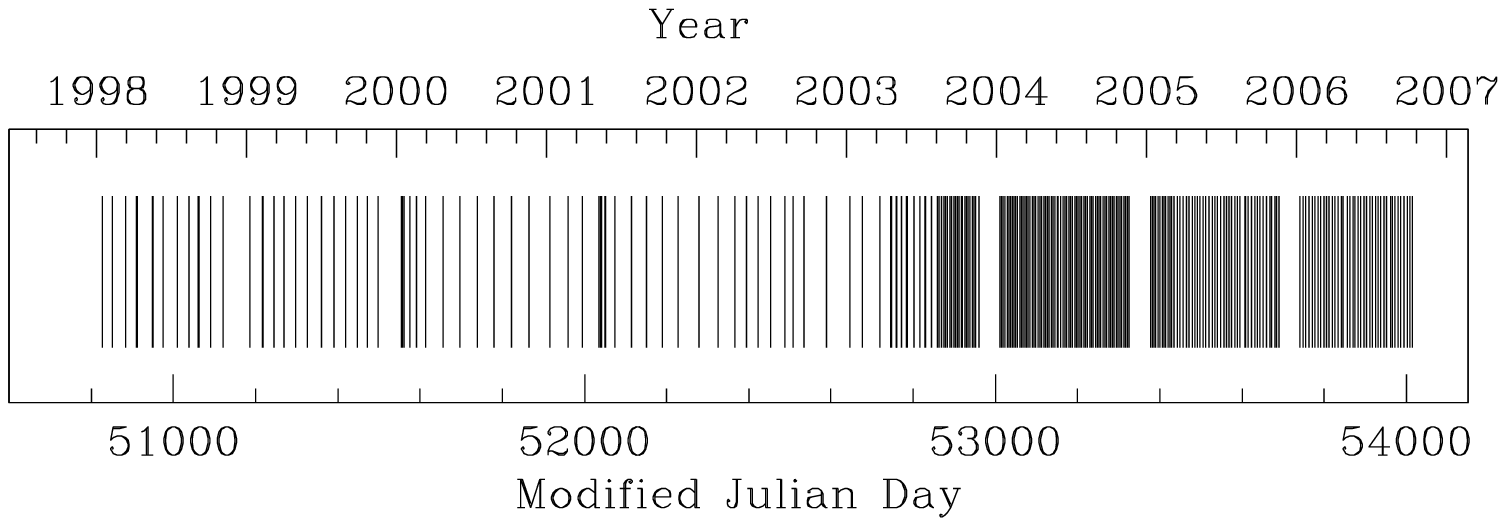}
\caption{Epochs of observations of \seven\ with {\em{RXTE}}.  Gaps near
the end/start of each year are due to Sun avoidance.  See
Table~\ref{tableobs1} for details.
\label{figureobs1}}
\end{figure}

\begin{figure}
\includegraphics{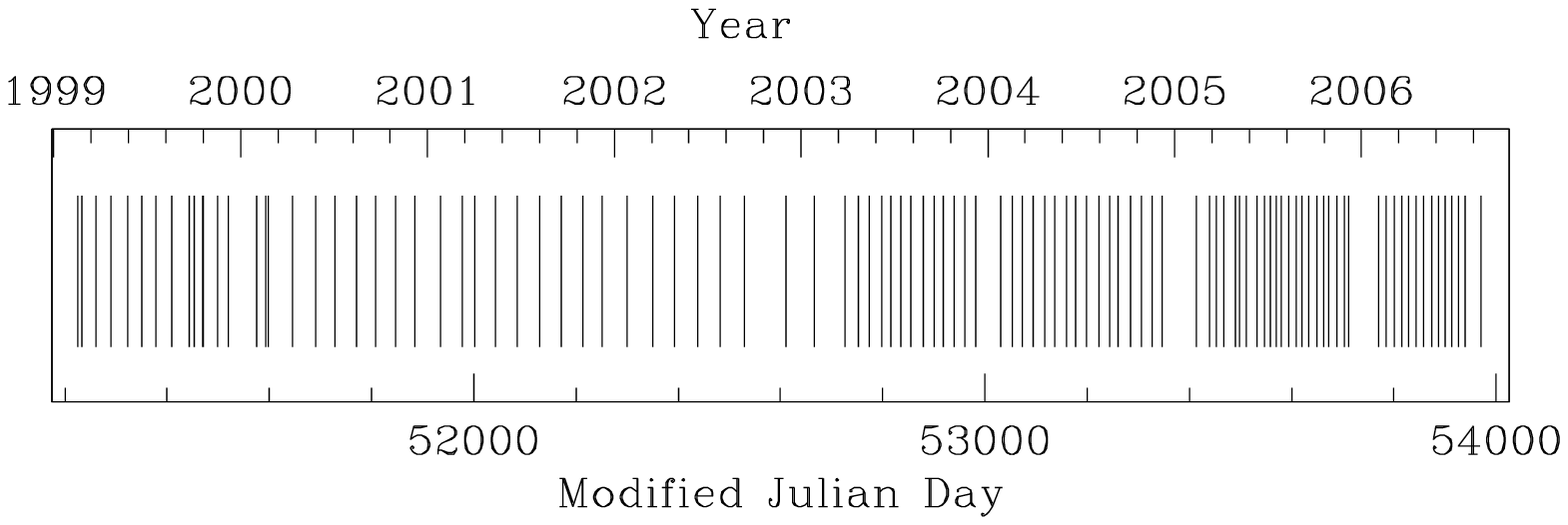}
\caption{Epochs of observations of \eight\ with {\em{RXTE}}.  Gaps near
the end/start of each year are due to Sun avoidance.  See 
Table~\ref{tableobs2} for details.
\label{figureobs2}}
\end{figure}

The results presented here were obtained using the PCA on board {\em{RXTE}}.
The PCA consists of an array of five collimated xenon/methane multi-anode
proportional counter units (PCUs) operating in the 2$-$60 keV range, with a
total effective area of approximately 6500~cm$^2$ and a field of view of
$\sim$1$^{\circ}$ FWHM \citep{jsg+96}. Our 294 observations of \seven\ and
our 136 observations of \eight\ are of various lengths (see
Tables~\ref{tableobs1} and~\ref{tableobs2}). Most were obtained over a
period of several years as part of a long-term monitoring program, but some
are isolated observations (see Figures~\ref{figureobs1}
and~\ref{figureobs2}).

For the monitoring, we used the {\tt GoodXenonwithPropane} data mode except during
Cycles~10 and~11 when we used the {\tt GoodXenon} mode.  Both data modes
record photon arrival times with 1-$\mu$s resolution and bin photon energies into
one of 256 channels. To maximize the signal-to-noise ratio, we analysed only
those events from the top xenon layer of each PCU.

\section{Phase-Coherent Timing}
\label{sec:timing}

Photon arrival times at each epoch were adjusted to the solar system barycenter.
Resulting arrival times were binned with 31.25-ms time resolution. In the
\seven\ timing analysis, we included only events in the energy range
2$-$6~keV, to maximize the signal-to-noise ratio of the pulse. Similarly,
for \eight\ we included events in the energy range 2$-$11~keV.

Each barycentric binned time series was folded using an ephemeris
determined iteratively by maintaining phase coherence as we describe below.
Resulting pulse profiles, with 64 phase bins, were cross-correlated in the
Fourier domain with a high signal-to-noise template created by adding
phase-aligned profiles from all observations. The cross-correlation returned
an average pulse time of arrival (TOA) for each observation corresponding to
a fixed pulse phase. The pulse phase $\phi$ at any time $t$ can usually be
expressed as a Taylor expansion,
\begin{equation}
\phi(t) = \phi_{0}(t_{0})+\nu_{0}(t-t_{0})+ \frac{1}{2}\dot{\nu_{0}}(t-t_{0})^{2} +\frac{1}{6}\ddot{\nu_{0}}(t-t_{0})^{3}+{\ldots},
\end{equation}
where $\nu$~$\equiv$~1/$P$ is the pulse frequency,
$\dot{\nu}$~$\equiv$~$d\nu$/$dt$, etc$.$, and subscript ``0'' denotes a
parameter evaluated at the reference epoch $t=t_0$. The TOAs were fitted to
the above polynomial using the pulsar timing software package
TEMPO\footnote{See http://www.atnf.csiro.au/research/pulsar/tempo.}. 

Note that we also searched for X-ray bursts in each 2--20~keV barycentered,
binned time series using the methods described in \citet{gkw04}, however no
bursts were found in any of our \seven\ or \eight\ data sets.

\subsection{Timing Results for \seven\ }
\label{sec:timing1708}

\begin{figure}
\centerline{\includegraphics[scale=0.65]{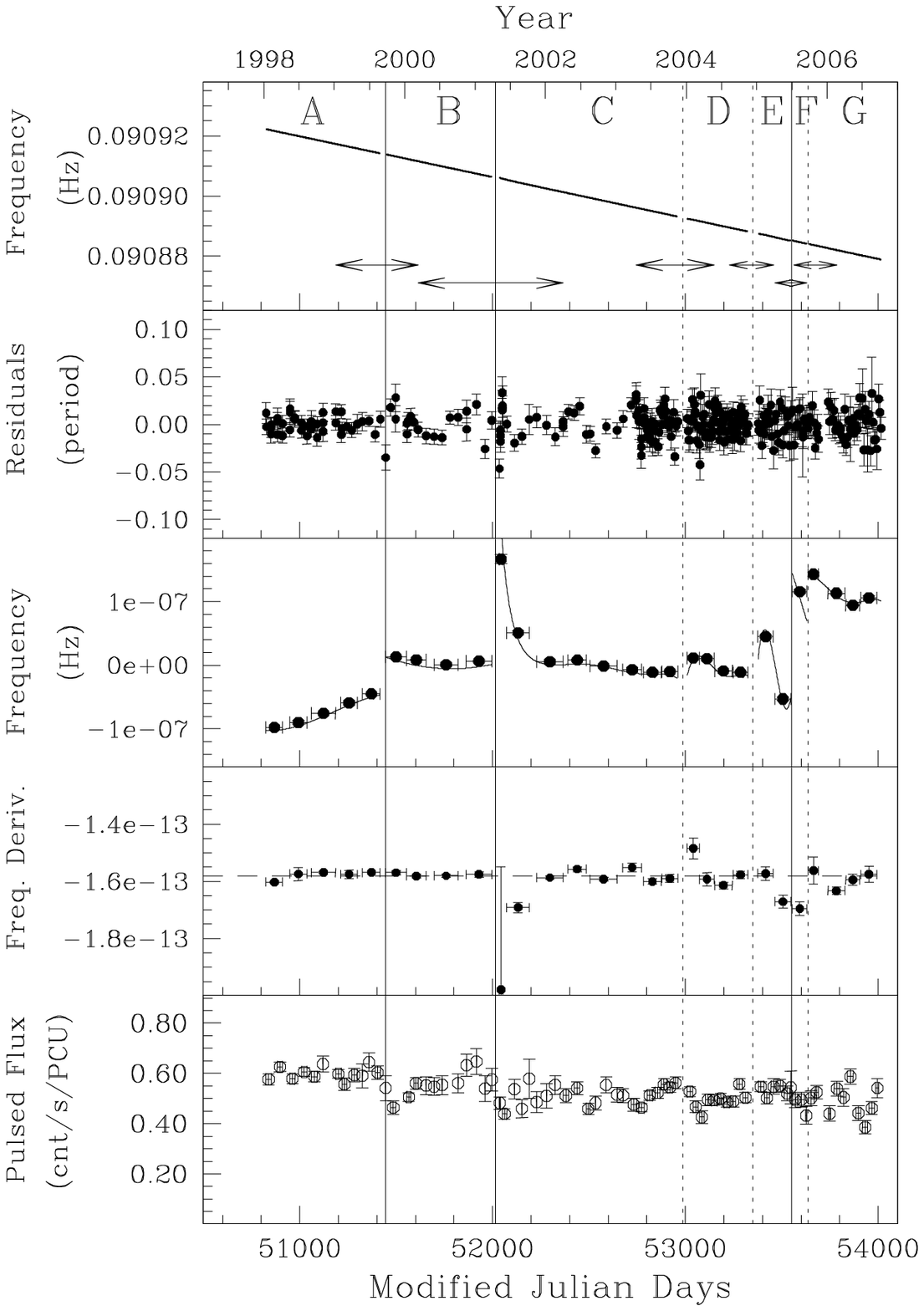}}
\caption {Spin and pulsed flux evolution in \seven.  Panels are described
from top to bottom.  Top: Frequency evolution, with inter-glitch intervals
indicated for correspondence with ephemerides given in
Table~\ref{table1708big}.  Arrows indicate intervals for which glitch
ephemerides were obtained (see Table~\ref{table1708glitches}).  Next:
residuals, after subtraction of the best-fit models given in 
Table~\ref{table1708big} (with arbitrary inter-interval phase offsets
subtracted). The increased scatter after MJD 52600 is due to a decrease in
typical integration time and an increase in monitoring frequency.  Next:
Solid curve: frequency evolution of the models shown in
Table~\ref{table1708big} after removal of the linear trend defined by the
frequency and frequency derivative from interval C as measured by fitting
only those parameters. Data points: measured frequencies in independent
sub-intervals after subtraction of the extrapolation of the same linear
trend. Next: Evolution of the frequency derivative in sub-intervals, when
fitting locally for only $\nu$ and $\dot{\nu}$.  Bottom: Pulsed flux in the
2--10 keV range.  All panels: Unambiguous glitch epochs are indicated with
solid vertical lines.  Candidate glitch epochs are indicated with dashed
vertical lines.
\label{figure1708big} }
\end{figure}

Figure \ref{figure1708big} and Table~\ref{table1708big} summarize our
results for \seven.  The pulsar's spin evolution can be characterized by
steady spin-down, punctuated by sudden episodes of spin-up, i.e., glitches,
in addition to candidate glitch events and apparently random noise. We
provide in Table~\ref{table1708big} pulse ephemerides for inter-glitch
ranges labelled as in the top panel of Figure~\ref{figure1708big}. Residuals
after subtraction of these models are shown in the next panel of
Figure~\ref{figure1708big}.  Overall the models describe the data well.
However, particularly when our timing precision was highest (i.e., before
2003), some low-level but significant deviations are seen on time scales of
weeks to months.  Their origin is unknown but is likely related to ``timing
noise,'' commonly seen in other AXPs \citep[e.g.,][]{kcs99,ggk+02} and
ubiquitously in radio pulsars \citep[e.g.,][]{antt94,hlk+04,lkgm05}. Note
that \cite{dkg07} performed simulations which showed that pulse profile
changes similar to those observed in this source (see
\S\ref{sec:profiles1708}) did not result in timing offsets significantly
larger than our reported TOA uncertainties. Hence, the features in the
timing residuals reported here are not a result of pulse profile changes.

\begin{figure}
\includegraphics[scale=0.7]{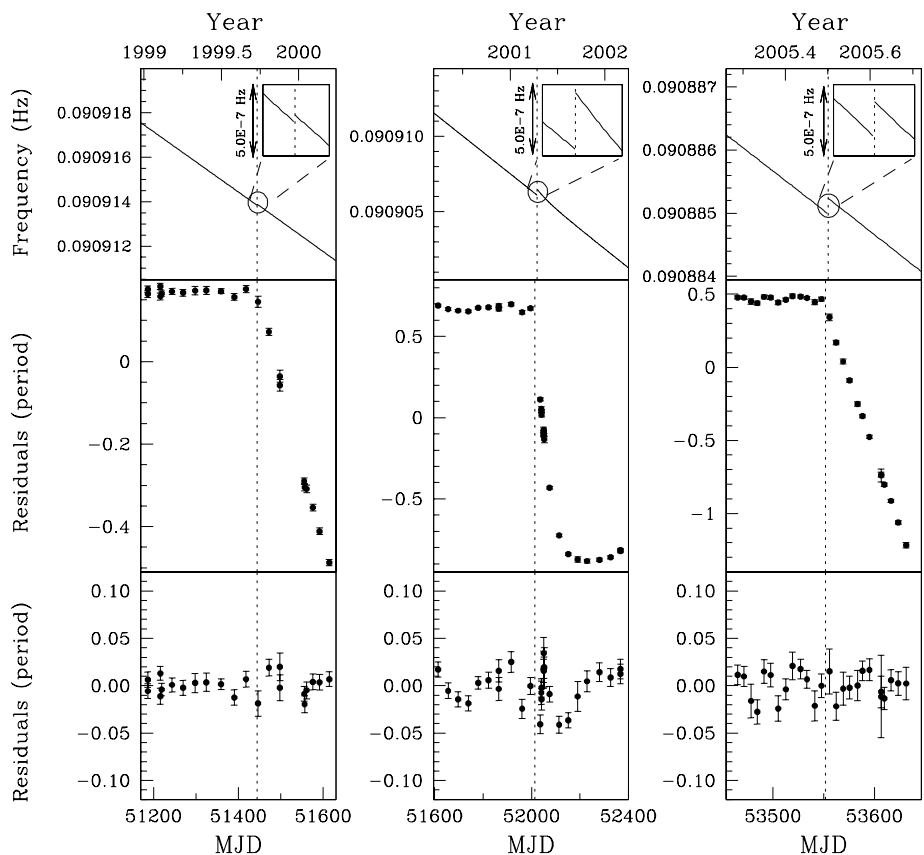}
\caption
{The three unambiguous glitches observed in \seven.  Top panels: Frequency
evolution around glitch as determined from ephemerides in
Table~\ref{table1708glitches}, with the blow-up inset displaying glitch
amplitude on a common scale for comparison.  Middle panels: Residuals after
subtraction of the best-fit pre-glitch ephemeris given in
Table~\ref{table1708glitches}. Bottom panels: Residuals after subtraction of
best-fit glitch models given in Table~\ref{table1708glitches}. All panels:
Dashed vertical lines indicate assumed glitch epochs.
\label{figure1708glitches}
}
\end{figure}

In addition to the two previously reported glitches (which we have
reanalysed, finding results consistent with those already in the literature;
see \citealt{klc00,kg03,dis+03}), we have identified a third unambiguous
glitch that occured near MJD 53551 (2005 June 30).  Note that the exact
glitch epoch is unknown due to our non-continuous monitoring; we report an
epoch for which the phase jump is zero. This is because a non-zero phase
jump at the time of the glitch would suggest an unphysically large torque on
the star. This third glitch had fractional frequency jump $\Delta\nu/\nu =
2.7 \times 10^{-6}$, and no obvious recovery.  This glitch amplitude is
intermediate between those of the previous two observed glitches, and the
lack of recovery is similar to what was seen in the first glitch, but in
marked contrast with the second glitch, as is clear from
Figure~\ref{figure1708glitches}.  A sudden change in post-glitch spin-down
rate for the third glitch is difficult to constrain, because of a possible
additional glitch that occured not long after, as we describe below. Indeed
glitch-induced long-term changes in $\dot{\nu}$ aside from that following
the first glitch, as described by \citet{klc00}, are difficult to identify
given the apparent timing noise processes. Table~\ref{table1708glitches}
summarizes the parameters of the three certain glitches of \seven, assuming
a glitch model consisting of a permanent change in $\nu$ and $\dot{\nu}$ and
a frequency change $\nu_d$ that decayed on a time scale of $\tau_d$, i.e.,
\begin{equation}
\nu = \nu_0(t) + \Delta\nu + \Delta\nu_{d}e^{-(t-t_g)/{\tau_d}} + \Delta\dot{\nu}~(t-t_g),
\label{eq:glitch}
\end{equation}
where $\nu_0(t)$ is the frequency evolution pre-glitch, $\Delta \nu$ is a
instantaneous frequency jump, $\Delta \nu_d$ is the post-glitch frequency
increase that decays exponentially on a time scale $\tau_d$, $t_g$ is the
glitch epoch, and $\Delta \dot{\nu}$ is the post-glitch change in the
long-term frequency derivative.

For the second glitch, residuals after subtraction of a simple glitch with
fractional exponential recovery have clear remaining trends, as is clear in
Figures \ref{figure1708big} (second panel) and \ref{figure1708glitches}
(bottom panel).  Systematic trends after simple glitch model subtraction
were also reported by \citet{wkt+04} for the 2002 glitch in 1E~2259+586.  We
also find this in the largest glitch in \eight\ (see
\S~\ref{sec:timing1841}).  \citet{wkt+04} showed that for 1E~2259+586, the
glitch fit was significantly improved by adding an exponential growth term.
We have tried fitting this model to the second glitch from \seven\ but find
no improvement, with the preferred growth term consistent with zero.

\begin{figure}
\includegraphics[scale=0.7]{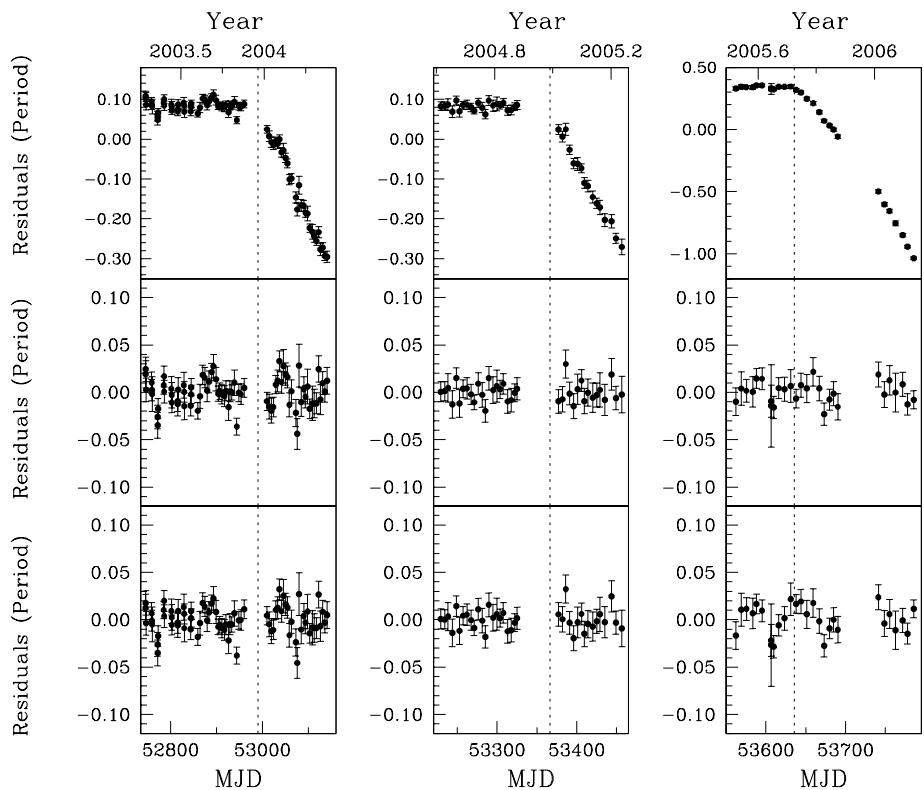}
\caption
{The three candidate glitches in \seven.  Top panels: Residuals after
subtraction of best-fit pre-`glitch' ephemeris given in 
Table~\ref{table1708cands}.  Middle panels: Residuals after subtraction of
best-fit glitch models given in Table~\ref{table1708cands}. Bottom panels:
Residuals after subtraction of best-fit alternative models given in
Table~\ref{table1708alt}. All panels: Dashed vertical lines indicate assumed
glitch epochs.
\label{figure1708cands}}
\end{figure}

In addition to the new certain glitch we report above, we find strong
evidence for an additional three glitches, each having fractional amplitude
similar to the first certain glitch seen in this source.  The properties of
these candidate glitches are summarized in Table~\ref{table1708cands}.
Timing residuals around the epochs of these glitches are shown in
Figure~\ref{figure1708cands}, in the top panel. Residuals following the
subtraction of a glitch model are shown in the middle panel of that Figure.
We refer to these as candidates only because a 4-th order polynomial fit to
the same data results in similar residuals (bottom panel of
Fig.~\ref{figure1708cands}; Table~\ref{table1708alt}) without the need to
invoke a sudden event.  The distinction between true glitches and timing
noise is often difficult to make for small-amplitude glitches, as discussed
by \citet{klc00}.  One way to distinguish, at least statistically, is that
apparent discontinuities attributable to timing noise should not have a
preferential direction, i.e., apparent spin-down `glitches' should be seen
too.  An examination of the frequency panel in Figure~\ref{figure1708big}
reveals apparent frequency jumps at the candidate glitch levels in both
directions, suggesting one or more of the candidates could indeed be timing
noise. Continued monitoring to acquire a larger database of such apparent
discontinuities will help clarify this issue.

Subsequent to our submission and posting of this paper, \cite{igz+07} posted
the results of a similar analysis of a subset of these same data. Some of
their results are consistent with ours however others differ. They reported
two glitches, the first of which corresponds to our second candidate glitch
(Table~\ref{table1708cands}). For that glitch, the reported fit parameters
are similar though not identical to ours.  Their second glitch corresponds
to our third glitch in Table~\ref{table1708glitches}. For that glitch, the
reported frequency jump at the glitch epoch was similar to ours but the jump
in frequency derivative was significantly different.  We find that this
difference is due to their inclusion of more post-glitch TOAs when fitting
the glitch. We did not include these TOAs because of a candidate event that
occurs shortly thereafter, but which Israel et al. did not report. In
addition to this difference, the frequency value reported in their
post-glitch ephemeris, 0.09088624(2)~Hz, is 42$\sigma$ away from the value
0.090885327(8)~Hz that we measure at the same epoch using our post-glitch
ephemeris. The numbers in parentheses are 1$\sigma$ uncertainties. We do not
understand this difference.

\subsection{Timing Results for \eight\ }
\label{sec:timing1841}

\begin{figure}
\centerline{\includegraphics[scale=0.65]{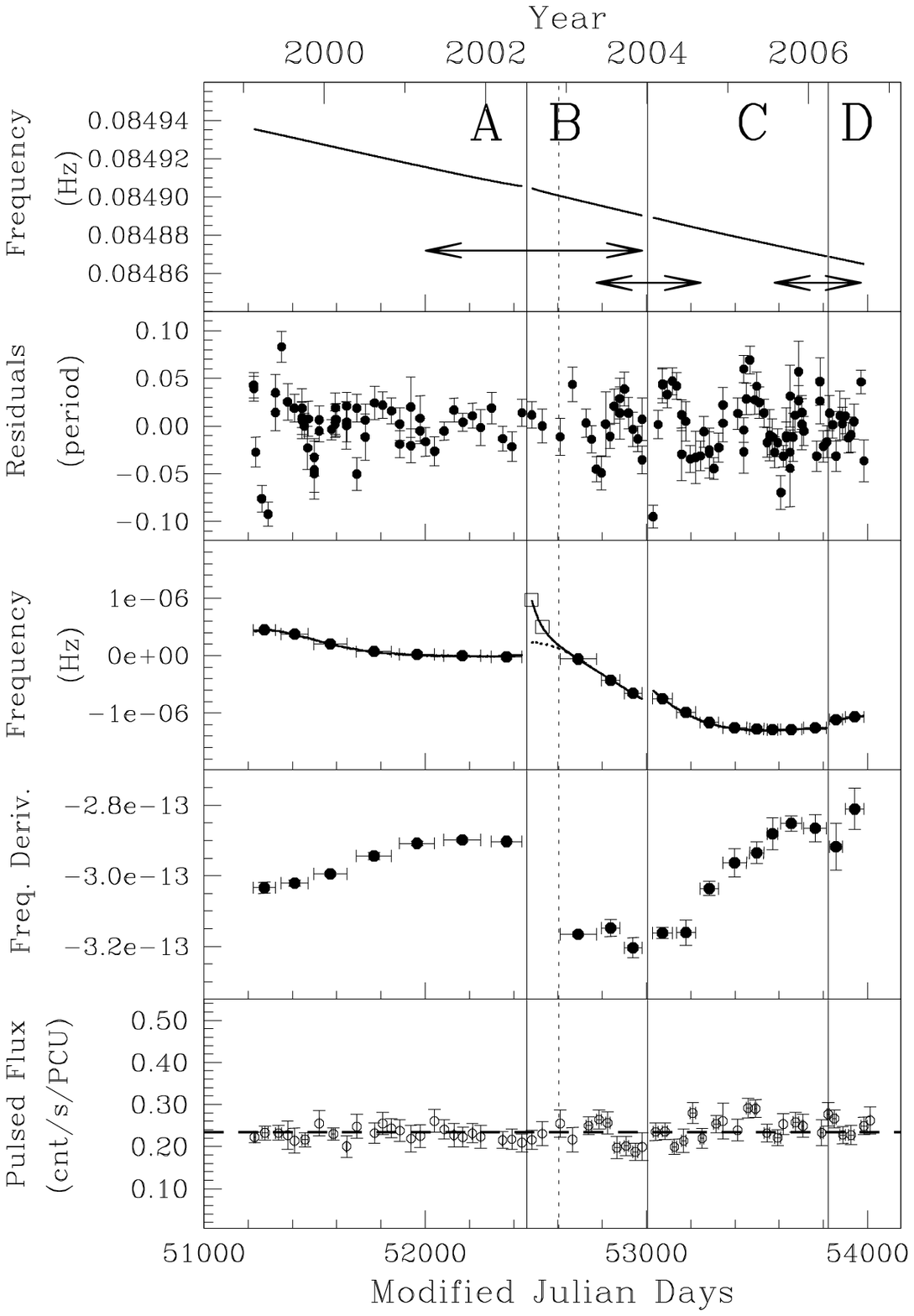}}
\caption {Spin and pulsed flux evolution in \eight.  Panels are
described from top to bottom.  Top: Frequency evolution, with inter-glitch
intervals indicated for correspondence with ephemerides given in
Table~\ref{table1841big}.  Arrows indicate intervals for which glitch
ephemerides were obtained (see Table~\ref{table1841glitches}). Next:
Residuals, after subtraction of the best-fit models given in
Table~\ref{table1841big}.  Next: Solid curve: frequency evolution of the
models shown in Table~\ref{table1841big} after removal of the linear trend
defined by the frequency and frequency derivative from the last year of data
before the first glitch, as measured by fitting only those parameters.
Dotted curve: alternate glitch recovery (see Section~\ref{sec:timing1841}
for details). Filled circles: Measured frequencies in independent
sub-intervals after subtraction of the extrapolation of the same linear
trend.  Unfilled squares: Epochs of the two immediate post-glitch
observations (too few for the measurement of an independent frequency but
crucial for the phase-coherent analysis).  Next: Evolution of the frequency
derivative in sub-intervals, when fitting locally for only $\nu$ and
$\dot{\nu}$.  Bottom: Pulsed flux in the 2--10 keV range.  All panels:
Glitch epochs are indicated with solid vertical lines.  The dashed vertical
line indicates the start of ephemeris B2, which does not include the two
immediate post-glitch observations (indicated with unfilled squares).
\label{figure1841big}}
\end{figure}

Figure~\ref{figure1841big} and Table~\ref{table1841big} summarize the
long-term timing behavior of \eight.  As for \seven, the spin evolution of
\eight\ is well characterized by regular spin-down punctuated by occasional
sudden spin-up events, plus timing noise.  Ephemerides in Table
\ref{table1841big} are given for the glitch-free intervals indicated in the
top panel of Figure~\ref{figure1841big}.  As for \seven, the long-term
timing residuals show some unmodelled trends whose origin is unknown. We
consider these trends timing noise, as did \citet{ggk+02} in their analysis
of $\sim$2~yr of data from this object.  Note that the ephemeris in Table
\ref{table1841big} labeled B2 is the same as that labeled B except for the
omission of data immediately post-glitch (see caption to
Fig.~\ref{figure1841big}).

\begin{figure}
\includegraphics[scale=0.7]{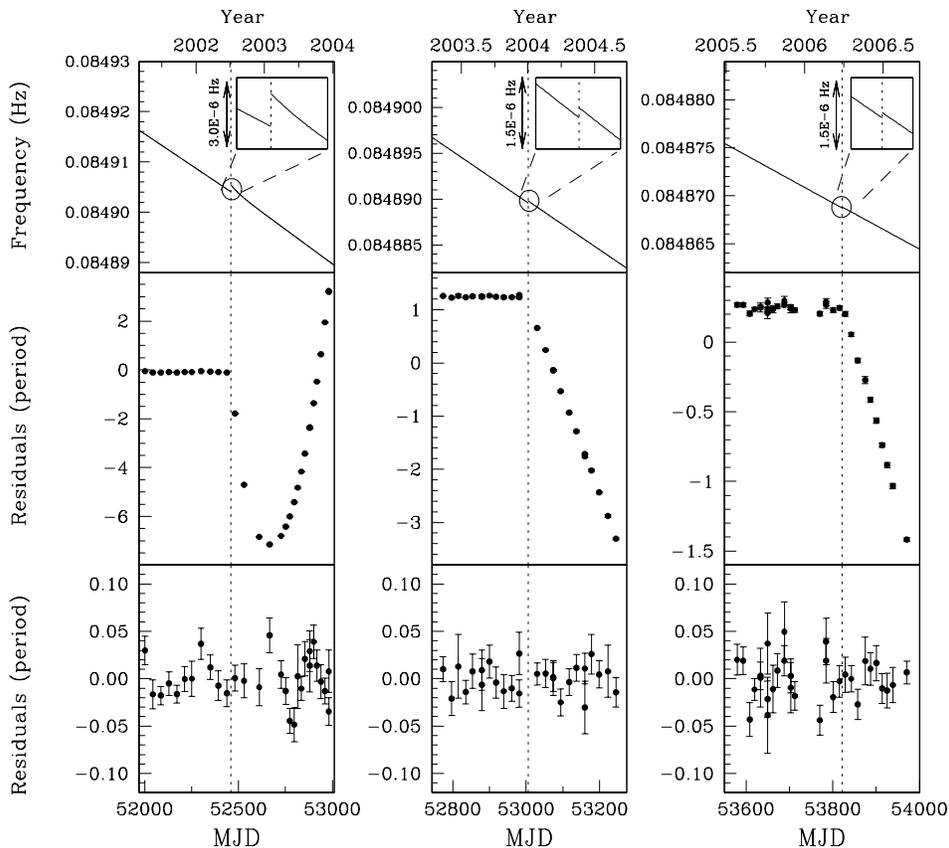}
\caption
{The three glitches in \eight.  Top panels: Frequency evolution around
glitches as determined from ephemerides in Table~\ref{table1841glitches},
with the blow-up inset displaying glitch amplitude. Middle panels: Residuals
after subtraction of the best-fit pre-glitch ephemeris given in
Table~\ref{table1841glitches}. Bottom panels: Residuals after subtraction of
best-fit glitch models given in Table~\ref{table1841glitches}. All panels:
Dashed vertical lines indicate assumed glitch epochs.
\label{figure1841glitches}}
\end{figure}

The frequency panel in Figure~\ref{figure1841big} and first panel in
Figure~\ref{figure1841glitches} make clear that \eight\ suffered a large
glitch, with significant recovery, near MJD 52460 (2002 July 5). This epoch
is estimated, as for all glitches reported in this paper, by taking the
epoch at which the phase jump is zero. Note that for this glitch there were
several such epochs and the one we are reporting gives the most conservative
frequency jump assuming an exponential recovery. The least conservative
possible frequency jump is $\sim$~50\% larger. The glitch fractional
amplitude was $\Delta\nu/\nu = 1.6\times10^{-5}$ (see
Table~\ref{table1841glitches}), among the largest yet seen from any neutron
star. A fraction $Q\equiv \Delta\nu_d/(\Delta\nu_d + \Delta\nu) = 0.64$ of
the glitch recovered on a time scale of 43~days. This glitch is thus similar
to the second certain glitch seen in \seven, and to the 2002 glitch in
1E~2259+586, which also showed significant recoveries on time scales of
weeks.  Also, like the second glitch of \seven, this large glitch in \eight\
is not well modelled by Equation~\ref{eq:glitch}, as is clear in the
residuals plot in Figure~\ref{figure1841glitches}.  Accompanying this
frequency glitch was a substantial long-term increase in the the magnitude
of $\dot{\nu}$, with fractional increase
$\Delta\dot{\nu}/\dot{\nu}=0.0959\pm 0.0007$. This is discussed further in
\S\ref{sec:disc_glitch}.

Because of the sparsity of the data around the glitch epoch, we found an
alternate ephemeris for the period of time covered by ephemeris~B
(Fig.~\ref{figure1841big}, Table~\ref{table1841big}). The fit parameters are
$\nu=0.0849041677(17)$~Hz, $\dot{\nu}=-2.852(3)\times10^{-13}$~Hz~s$^{-1}$,
$\ddot{\nu}=-2.47(8)\times10^{-21}$~Hz~s$^{-2}$, and
$d^{3}\nu/dt^{3}=8.8(7)\times10^{-29}$~Hz~s$^{-3}$ at the reported glitch
epoch MJD 52464.00448, with RMS phase residual of 0.019. This ephemeris
disagrees with ephemeris B in the shape of the recovery (see dotted curve in
Panel~3 of Fig.~\ref{figure1841big}) but agrees with it after the end of the
recovery. Using the parameters of this alternate ephemeris, the change in
$\nu$ at the glitch epoch would be $2.20(3)\times10^{-7}$~Hz much smaller
than the one reported in Table~\ref{table1841glitches}. However, we hesitate
to interpret the glitch using this ephemeris because of the very unusual and
unique shape of the recovery it predicts. Note that this alternate ephemeris
also shows a long-term increase in the magnitude of $\dot{\nu}$ after the
glitch.

We also report the detection of two additional, smaller glitches, as
summarized in Table~\ref{table1841glitches} and displayed in
Figures~\ref{figure1841big} and \ref{figure1841glitches}.  Neither glitch
displays significant recovery, and both are well modelled by a simple
permanent frequency jump.

\section{Pulse Profile Changes}
\label{sec:profiles}

Another interesting AXP property we can study thanks to {\it RXTE}
monitoring is the evolution of the pulse profile. We performed a pulse
profile analysis on each AXP using FTOOLS version
5.3.1\footnote{http://heasarc.gsfc.nasa.gov/ftools}. We used the following
steps: for each observation, we ran the FTOOL \verb|make_se| to combine the
GoodXenon files.  We then used the FTOOL \verb|fasebin| to make a
phase-resolved spectrum of the entire observation with 64 phase bins across
the profile.  When we ran \verb|fasebin|, we selected layer~1 of the
detector, disregarded the propane photons, and included the photons from
PCUS~1,~2,~3, and~4.  We omitted PCU~0, for which an independent analysis of
AXP 4U 0142+61 revealed spectral modeling irregularities \citep{dkg07}.
\verb|fasebin| also took care of barycentering the data. For each
observation, we then used \verb|seextrct| to make a phase-averaged spectrum
for the same set of detector layers and PCUs. The phase-averaged spectrum
was then used by the perl script \verb|pcarsp| to make a response matrix.

We loaded the phase-resolved spectra and the response matrices into the
X-ray Spectral Fitting Package (XSPEC\footnote{http://xspec.gsfc.nasa.gov
Version: 11.3.1}) and selected photons belonging to three energy bands:
2$-$10, 2$-$4, and~4$-$10~keV. Using XSPEC, we extracted a count-rate pulse
profile for each of the energy bands. The profiles included XSPEC-obtained
1$\sigma$ error bars on each of the phase bins. To obtain a pulse profile in
units of count rate per PCU, we divided the overall profile by a PCU
coverage factor that took into account the amount of time each PCU was on.

We then aligned the 64-bin profiles with a high signal-to-noise template
using a similar cross-correlation procedure to the one used in the timing
analysis.  Then, for each glitch-free interval, we summed the aligned
profiles, subtracted the DC component, and scaled the resulting profile so
that the value of the highest bin is unity and the lowest point is zero.

\subsection{Profile Analysis results of \seven\ }
\label{sec:profiles1708}

\begin{figure}
\includegraphics{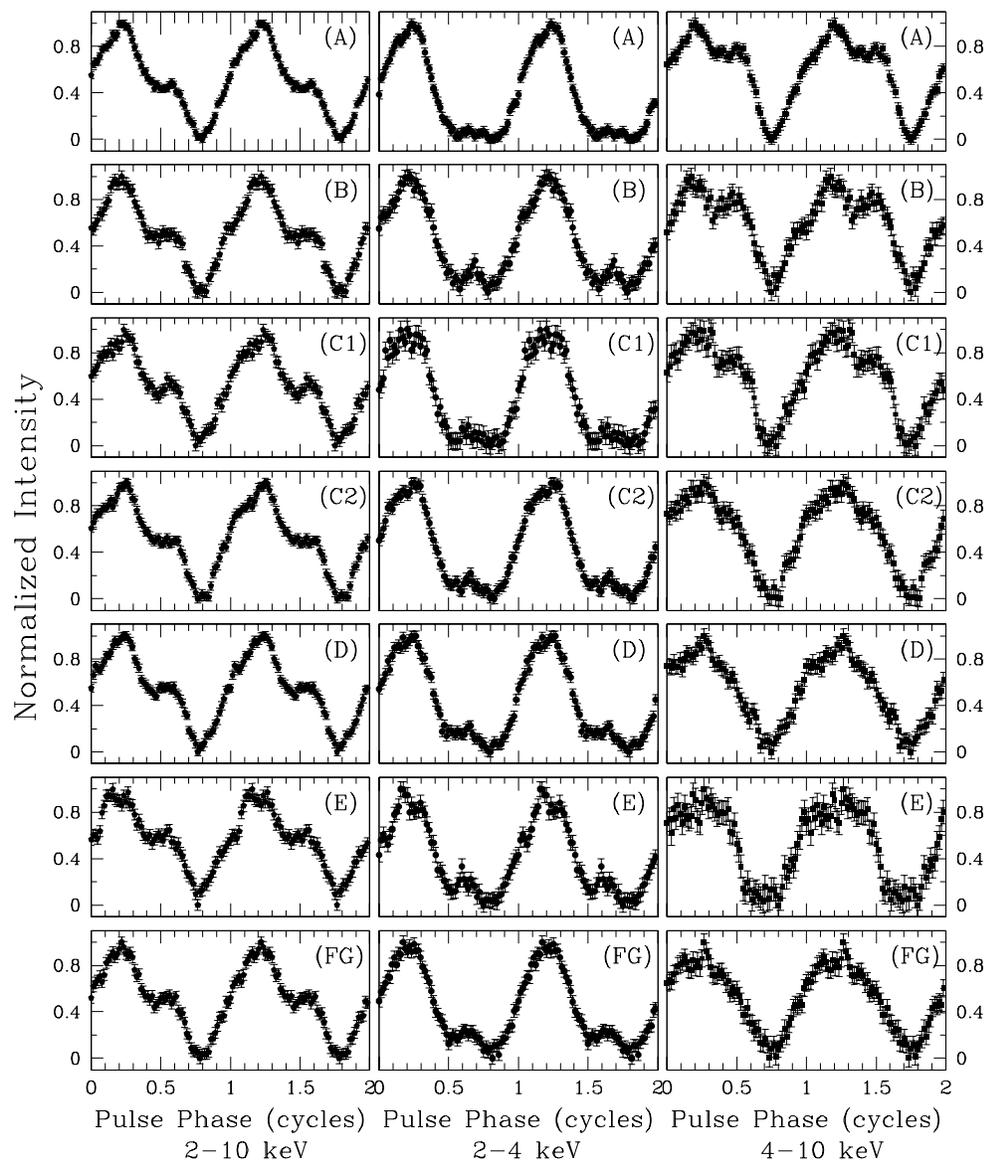}
\caption
{Normalized pulse profiles in three energy bands for \seven\ for the seven
glitch-free intervals (with corresponding labels at the top right) defined
in the top panel of Figure \protect\ref{figure1708big}. Different data
qualities within an energy range are due to different net exposure times.
Two cycles are shown for each profile for clarity.
\label{figure1708profiles}}
\end{figure}

Average profiles for \seven\ in the three energy bands are presented in
Figure~\ref{figure1708profiles}. In a given band, the different profile
qualities are due to different net exposure times. Energy dependence is
clearly visible to the eye as well as small fluctuations. For example, in
the 2--4~keV band, the small peak off the main pulse has clearly fluctuating
intensity.  This small peak gets larger at higher energy, as seen in the
2--10~keV band.  In the 4--10~keV band, the smaller peak seems to blend with
the main low-energy peak to yield a broad single peak structure, although
fluctuations in that structure are apparent.

\begin{figure}
\includegraphics[scale=1.2]{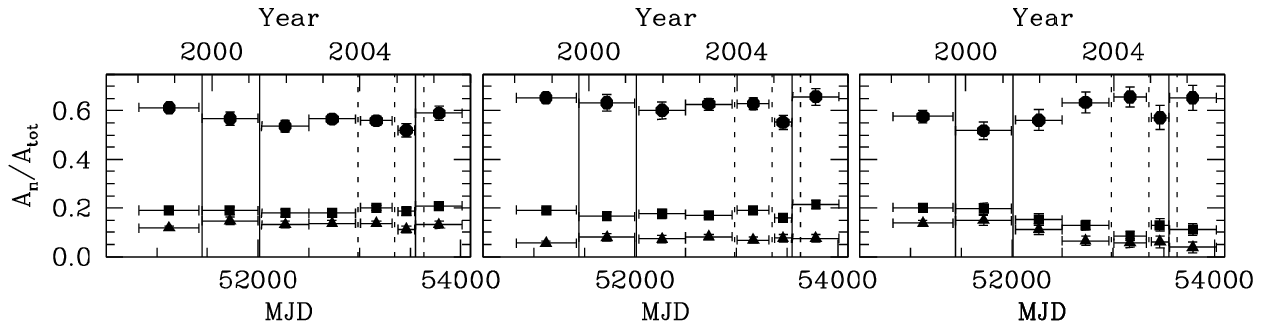}
\caption
{Left panel: Time evolution of the ratio of the power in the $n$th harmonic
to the total power in the 2--10 keV pulse profile of \seven. Circles
represent $n=1$, squares $n=2$, and triangles $n=3$.  Solid vertical lines
indicate epochs of glitches; dashed vertical lines are epochs of candidate
glitches. Middle panel: Same as left panel but for 2--4~keV. Right panel:
Same as middle but for 4--10~keV.
\label{figure1708harm1}}
\end{figure}

To study these fluctuations quantitatively, we subjected each profile to a
Fourier analysis.  Figure~\ref{figure1708harm1} shows the evolution of the
first three profile harmonics with time.  Although there are hints of
variation in all energy bands, only variations in the hard 4--10~keV band
are statistically significant (as determined by the $\chi^2$ statistic from
a fit to a constant value); the decline of the second and third harmonics in
the hard band have probabilities of 0.0007\% and 0.0012\%, respectively, of
being due to chance.  Thus in the hard band the profile is certainly
becoming more sinusoidal, in agreement with what is inferred by eye.

\begin{figure}
\includegraphics{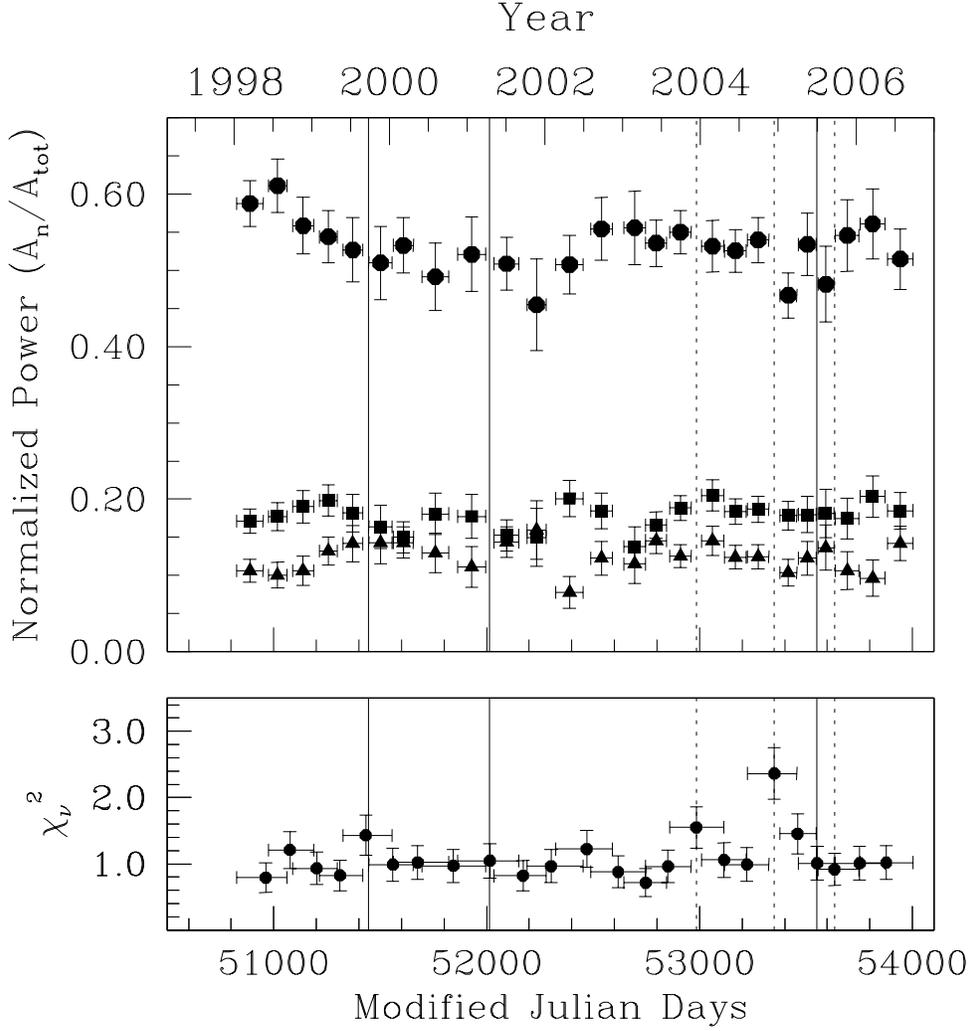}
\caption
{Top panel: Time evolution of the ratio of the power in the $n$th harmonic
to the total power in the 2--10 keV pulse profile of \seven.  The circles
represent $n=1$, squares $n=2$, and triangles $n=3$.  Solid and dashed
vertical lines indicate epochs of glitches and candidate glitches,
respectively.  The probability that the observed fluctuations are due to
random noise are 68\%, 97\% and 69\% for $n=1,2,3$, respectively. Bottom
panel: reduced $\chi^2$ per degree of freedom for successive profile
differences (see text \S\ref{sec:profiles1708} for details).
\label{figure1708harmonics}}
\end{figure}

The above analysis shows that the profile is changing, but not whether these
changes are truly correlated with the glitch epochs, since changes could be
occuring throughout.  To search for pulse profile changes correlated with
glitch epochs, as were claimed by \citet{dis+03}, we divided glitch-free
intervals into several sub-intervals (typically of duration $\sim$30 days)
for which independent profiles were created. The number of sub-intervals was
chosen by trading off signal-to-noise ratio for time resolution. These
sub-interval profiles were then Fourier analysed.  The evolution of the
Fourier powers in the first three harmonics in the 2--10~keV profile are
shown in the top panel of Figure~\ref{figure1708harmonics}.

To determine whether the apparent fluctuations are statistically
significant, we fit a constant value to each time series and from the
$\chi^2$ of the best fit, found that the probabilities of the fluctuations
being due to random noise are 68, 96 and 69\% for $n=1,2,3$, respectively.
This analysis thus shows no evidence for profile changes associated with the
glitch epochs, including the second glitch.  However the reduced
signal-to-noise ratios in the sub-interval average profiles, required for
interesting time resolution, makes us insensitive to subtle profile changes.
To search for glitch-correlated pulse profile changes in a different way,
for each sub-interval we calculated the reduced $\chi^2$ of the difference
between that sub-interval's average profile and the previous one.  The time
series of these $\chi^2$ values is shown in the bottom panel of
Figure~\ref{figure1708harmonics}.  There is clearly no evidence for any
profile change at the second glitch, or at the third certain glitch.  There
is some hint of profile changes at the first and second candidate glitches,
however a K-S test shows that our $\chi^2$ values as a group have a
probability of 39\% of originating from $\chi^2$ distribution.
Interestingly though, the probability of the single high $\chi^2$ value we
measure at the second candidate glitch occuring randomly is only $1.0\times
10^{-6}$; that at the first glitch is 1.7\% and at the first glitch
candidate is 0.4\%.  Thus we do find possible evidence in this analysis for
glitch-correlated pulse profile changes, though the best evidence for
significant changes occurs only at two candidate glitches, i.e., the lowest
amplitude events.

\subsection{Profile Analysis results of \eight\ }
\label{sec:profiles1841}

\begin{figure}
\includegraphics{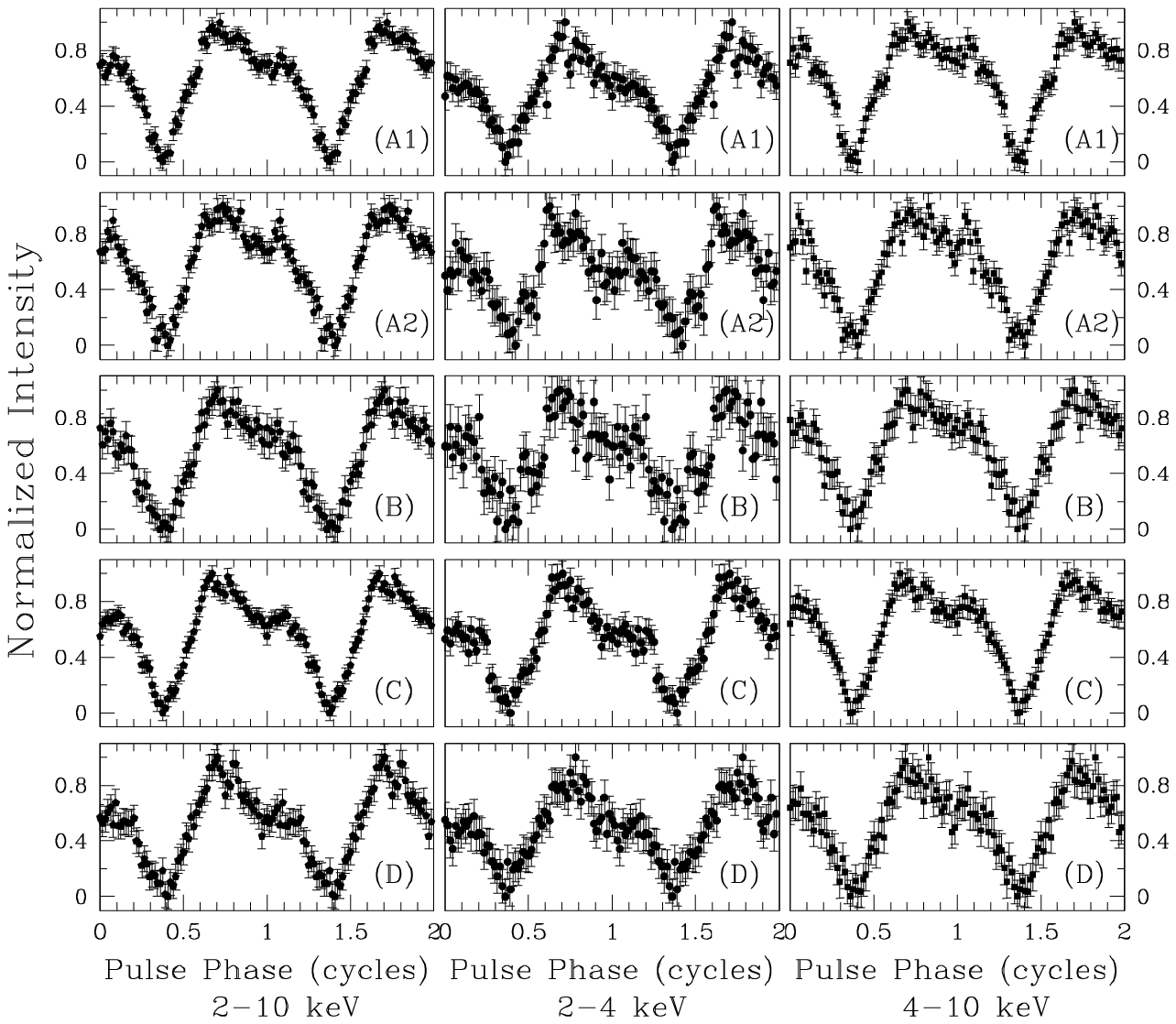}
\caption 
{Normalized pulse profiles in three energy bands for \eight\ for the five
glitch-free intervals defined in the top panel of
Figure~\protect\ref{figure1841big}.  Different data qualities in each energy
range are due to different net exposure times.  Two cycles are shown for
each profile for clarity.  
\label{figure1841profiles}}
\end{figure}

Summed profiles for \eight\ in three energy bands for the five glitch-free
intervals defined in the top panel of Figure~\ref{figure1841big} are shown
in Figure~\ref{figure1841profiles}.  As for \seven, some low-level profile
fluctuations are suggested, particularly in the relative amplitude of the
leading and trailing sides of the large single peak in the 2--10 keV band
(though clearly this peak could also be considered the blend of two or more
adjacent peaks).

\begin{figure}
\includegraphics[scale=1.2]{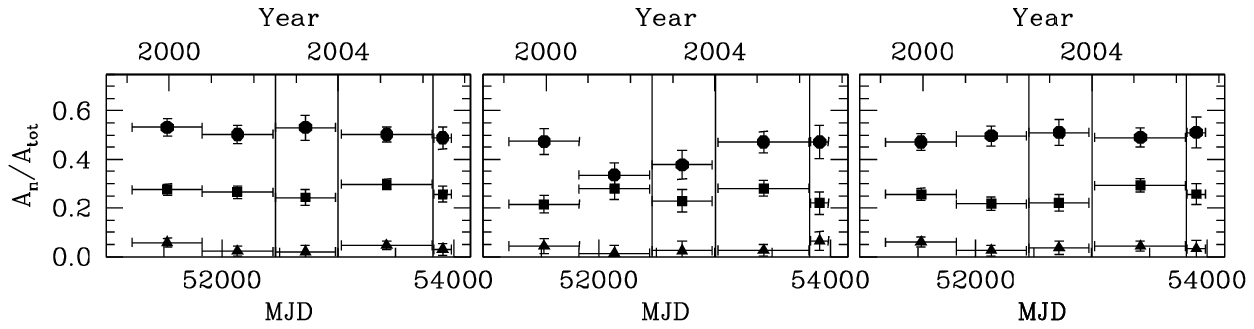}
\caption
{Left panel: Time evolution of the ratio of the power in the $n$th harmonic
to the total power in the 2--10 keV pulse profile of \eight. Circles
represent $n=1$, squares $n=2$, and triangles $n=3$.  Solid vertical lines
indicate epochs of glitches. Middle panel: Same as left panel but for
2--4~keV. Right panel: Same as middle but for 4--10~keV.
\label{figure1841harm1}}
\end{figure}

As for \seven, we quantify the profile fluctuations of \eight\ via Fourier
analysis.  Figure~\ref{figure1841harm1} shows the evolution of the first
three profile harmonics with time in each energy band.  Interestingly, in
contrast to \seven, here the profile changes are most prominent in the soft
2--4~keV band, in which the fraction of power in the fundamental of the
profile in interval A2 (see Fig.~\ref{figure1841big}) decreased, then slowly
relaxed back to the previous range.  However, a $\chi^2$ test shows the
probability of this behavior being due to random noise is 18\%, too large to
exclude.

\begin{figure}
\includegraphics{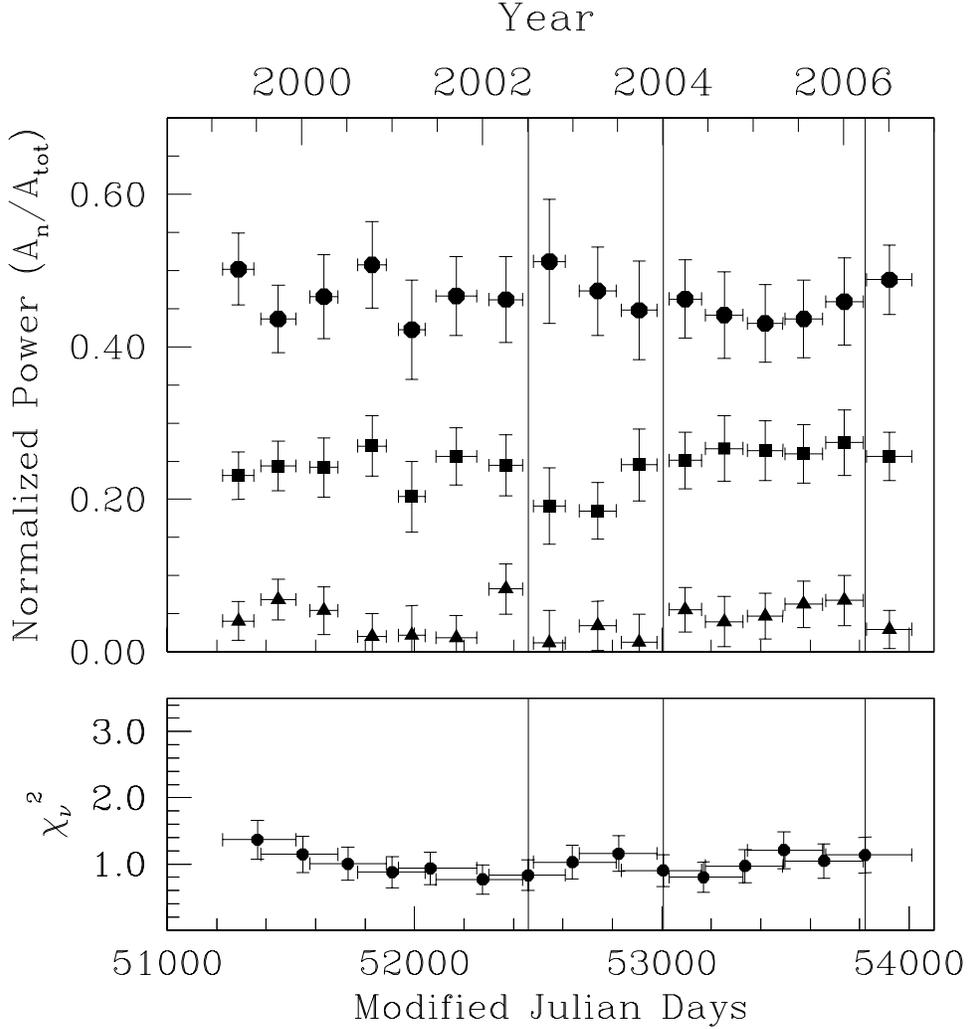}
\caption
{Top panel: Time evolution of the ratio of the power in the $n$th harmonic
to the total power in the 2--10 keV pulse profile of \eight. Circles
represent $n=1$, squares $n=2$, and triangles $n=3$.  Solid vertical lines
indicate epochs of glitches. The probabilities that the observed
fluctuations arise from random noise are 99\%, 97\% and 96\% for $n=1,2,3$,
respectively. Bottom panel: reduced $\chi^2$ per degree of freedom for
successive profile differences (see text \S\ref{sec:profiles1841} for
details).
\label{figure1841harmonics}}
\end{figure}

To look for pulse profile changes correlated with glitches, again,
sub-intervals within glitch-free intervals were chosen and summed profiles
computed and Fourier analysed. The evolution of the first three harmonics is
shown in the top panel of Figure~\ref{figure1841harmonics}.  Fluctuations
are apparent although none is particularly remarkable at any of the glitch
epochs, including the first and largest, and the time series for $n=1,2,3$
are all consistent with a constant value.  This argues again against
correlated profile and timing anomalies in this source thus far.  As a
confirmation, as for \seven, a difference profile was calculated for each
sub-interval by subtracting that interval's profile from the preceding one.
The $\chi^2$ values of these difference profiles are shown in the bottom of
Figure~\ref{figure1841harmonics}; no significant features are present.

\section{Pulsed Flux Time Series}
\label{sec:flux}

{\it RXTE} monitoring also allows the study of the evolution of the pulsed
flux of these sources.  To obtain a pulsed flux time series for \seven\ and
\eight, we did the following.  First, for each observation,
we used a procedure similar to that described in \S\ref{sec:profiles} 
to make a count rate per PCU pulse profile (with 64 phase bins across the
profile and excluding PCU~0) in the energy range 2--10~keV. The profiles
included XSPEC-determined 1$\sigma$ error bars on the flux value in each of
the phase bins.

The pulsed flux for each of the profiles was
calculated using the following RMS formula:
\begin{equation}
F = \sqrt{2 {\sum_{k=1}^{n}} (({a_k}^2+{b_k}^2)-({\sigma_{a_k}}^2+{\sigma_{b_k}}^2    ))} ,
\label{eq:pf}
\end{equation}
where $a_k$ is the $k^{\rm{th}}$ even Fourier component defined as $a_k =
\frac{1}{N} {\sum_{i=1}^{N}} {p_i} \cos {(2\pi k i/N})$, ${\sigma_{a_k}}^2$
is the uncertainty of $a_k$, $b_k$ is the odd $k^{\rm{th}}$ Fourier
component defined as $b_k = \frac{1}{N} {\sum_{i=1}^{N}} {p_i} \sin {(2\pi k
i/N})$, ${\sigma_{b_k}}^2$ is the uncertainty of $b_k$, $i$ refers to the
phase bin, $N$ is the total number of phase bins, $p_i$ is the count rate in
the $i^{\rm{th}}$ phase bin of the pulse profile, and $n$ is the maximum
number of Fourier harmonics to be taken into account. We used $n=6$ for both
\seven\ and \eight.

Our method for estimating the pulsed flux $F$ is equivalent to the simple
RMS formula $F =
\frac{1}{\sqrt{N}}\sqrt{{\sum_{i=1}^{N}}({p_i}-{\overline{p}})^2}$ (where
$p_i$ is the count rate in the $i^{\rm{th}}$ phase bin of the pulse profile
and $\overline{p}$ is the average count rate), except that we have
subtracted the variances (to eliminate the upward statistical bias) and only
included the statistically significant Fourier components. For a detailed
discussion on pulsed flux estimates, see Archibald, Dib \& Kaspi (in prep.).

\subsection{Pulsed Flux Time Series for \seven}
\label{sec:flux1708}

\begin{figure}
\includegraphics{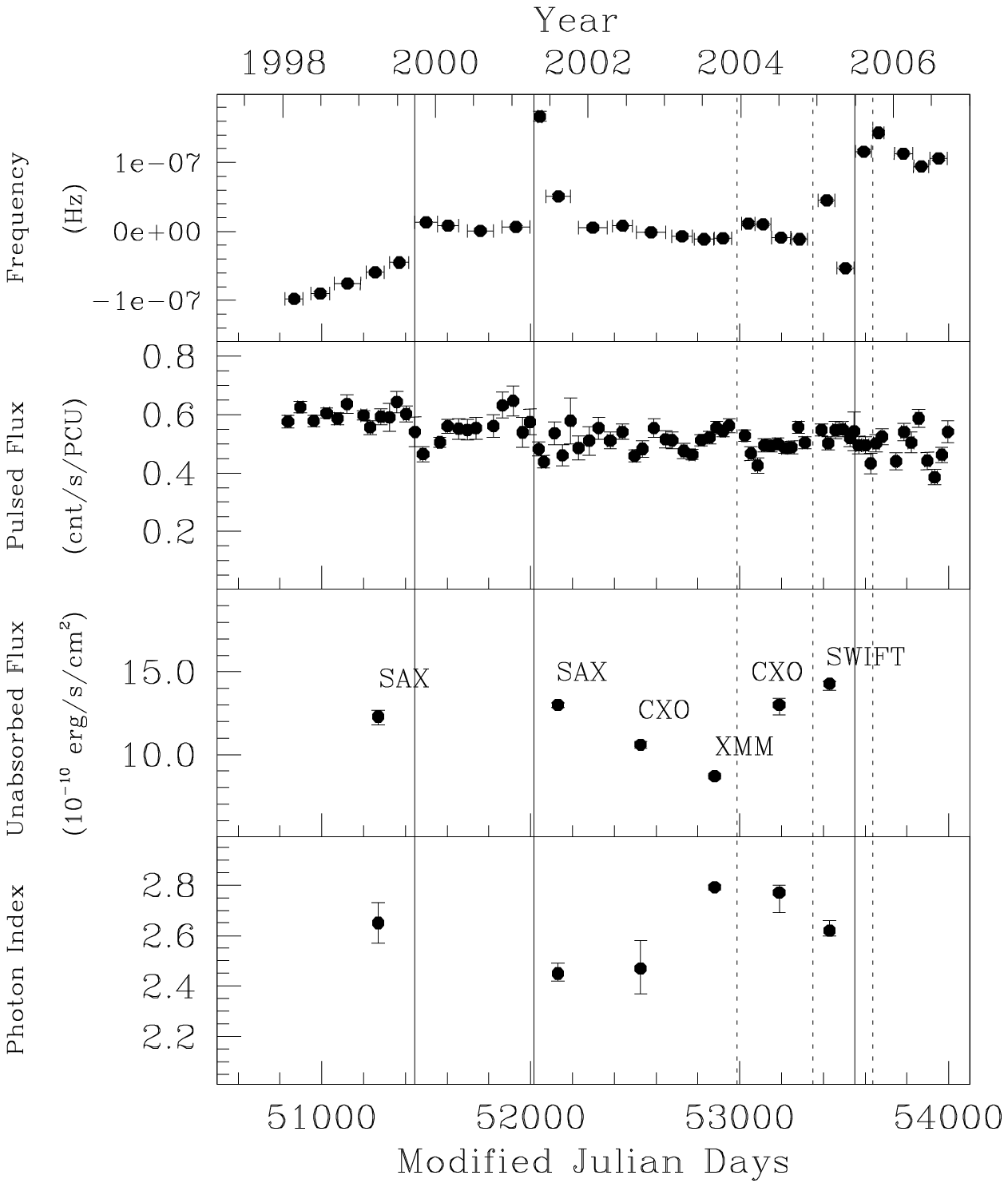}
\caption {
Frequency, pulsed flux, reported total unabsorbed flux, and reported photon
index as a function of time for \seven.  Frequency and pulsed flux data are
identical to those shown in Figure~\ref{figure1708big}.  Solid and dashed
vertical lines indicate epochs of glitches and glitch candidates,
respectively.  Unabsorbed phase-averaged 0.5--10 keV fluxes and photon
indexes are from \protect\citet{roz+05} and \protect\citet{cri+07}, and are
labelled by observing telescope.  That the pulsed flux remains relatively
constant while the phase-averaged flux appears to vary by nearly a factor of
two (albeit as measured by different instruments) suggests a strong
anti-correlation between total flux and pulsed fraction. 
\label{figure1708flux} }
\end{figure}

Our pulsed flux time series for \seven\ is shown in
Figure~\ref{figure1708big} (bottom panel) and again in
Figure~\ref{figure1708flux}.  Each data point represents the average of
pulsed fluxes measured over $\sim$1~month.  There appear to be frequent
low-level pulse flux variations in this source.  Although our error
estimates on the pulsed fluxes include only statistical uncertainty (i.e.,
we have made no effort to estimate systematic uncertainties), we are given
confidence that the fluctuations seen e.g., near MJD 52000 are real, given
how stable the pulsed flux of \eight\ is in the same time interval (see
\S\ref{sec:flux1841}).

There are no large increases in pulsed flux following any of the glitches,
unlike what was seen following the 2002 glitch of AXP 1E~2259+586
\citep{kgw+03,wkt+04}.  However, there is a possible pulsed flux enhancement
prior to the second glitch, and a dip following it.  Given this and the lack
of clearly associated pulse profile changes coincident with glitch epochs
(see \S\ref{sec:profiles1708}), the glitches of \seven\ appear to be
``quiet,'' in the sense that they seem unaccompanied by significant
pulsed radiative change.  This is discussed further in \S\ref{sec:discussion}.

Figure~\ref{figure1708flux} shows phase-averaged fluxes in the 0.5--10~keV
band as measured using a variety of focussing X-ray telescopes
\citep{roz+05,cri+07}.  Interestingly, while the reported phase-averaged
flux varies considerably (by a factor of $\sim$1.6), and in concert with the
photon index as measured in the conventionally used blackbody/power-law
spectral model, the 2--10~keV pulsed flux remains relatively constant.  This
suggests that the pulsed fraction of \seven\ is precisely anti-correlated
with total flux, in such a way as to keep the pulsed flux near constant.
This is discussed further in \S\ref{sec:discussion}.

The origin of the apparent low-level pulsed flux variations is not clear,
given the apparent lack of correlation with the phase-averaged flux.  As
shown by Archibald et al. (in prep.), a changing pulse profile can affect an
RMS-based pulsed flux estimator such as that in Equation~\ref{eq:pf}. To
verify that our measured pulsed fluxes were not influenced by the changing
pulse profile of \seven\ (see \S\ref{sec:profiles}), we also found the
pulsed flux using an estimator based on the area under the profile (after
baseline subtraction), which is, by definition, insensitive to pulse profile
changes.  With this method, we obtained qualitatively similar results for
the pulsed fluxes.

\subsection{Pulsed Flux Time Series for \eight}
\label{sec:flux1841}

Our pulsed flux time series for \eight\ is shown in the bottom panel of
Figure~\ref{figure1841big}.  Each data point represents the average of
pulsed fluxes measured over $\sim$1~month. Note the increased scatter after
MJD 52700 is due to decreased effective integration time, a result of the
reduction in the average number of operational PCUs. The measured pulsed
fluxes are consistent with being constant, with their probability of being
due to random fluctuations 52\%. There is no evidence for any pulsed flux
change at the glitch epochs. Thus the glitches of \eight\ appear to be
``quiet,'' at least in pulsed flux, on time scales comparable to or longer
than our sampling time.

\section{Discussion}
\label{sec:discussion}

\subsection{AXP Glitches}
\label{sec:disc_glitch}

We have now observed a sufficiently large sample of AXP glitches that we can
make meaningful phenomenological comparisons with glitches in radio pulsars,
a much better studied phenomenon.  Detection of systematic differences in
AXP and radio pulsar glitch properties would be interesting as it could
signal structural differences between magnetars and conventional radio
pulsars.

\begin{figure}
\includegraphics[scale=0.7]{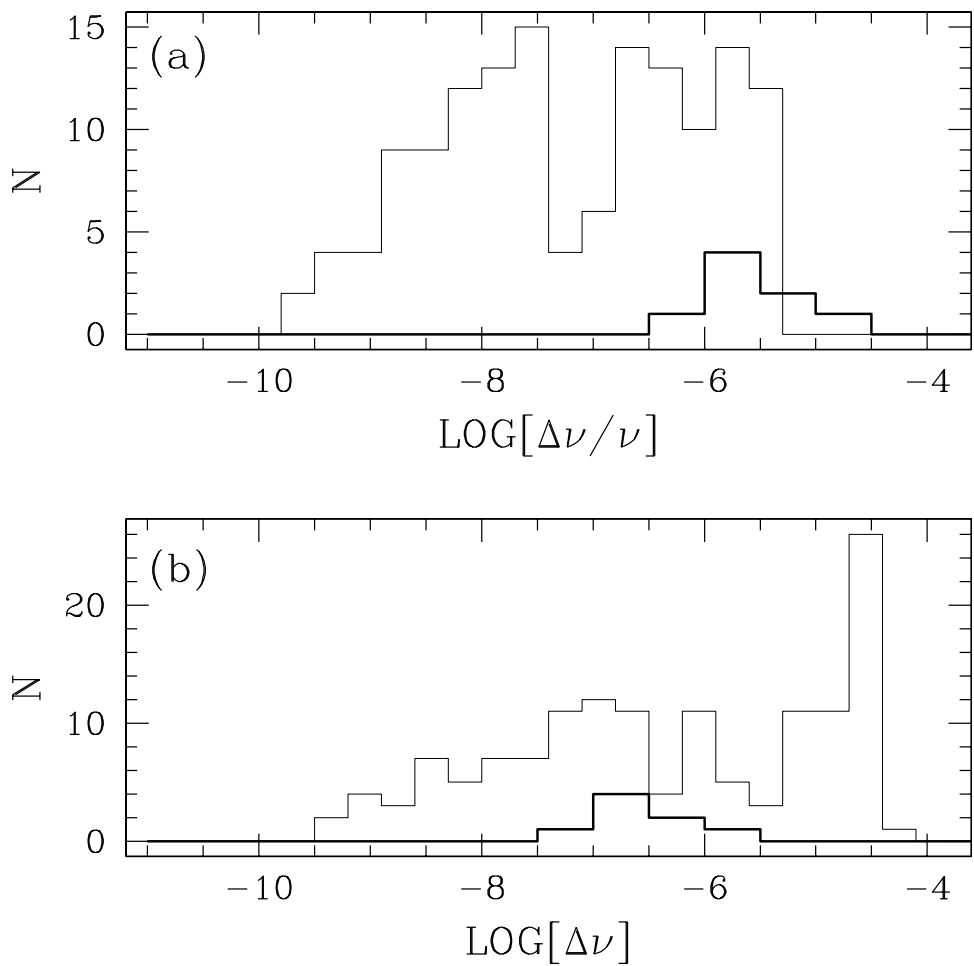}
\caption{
Amplitude distribution of AXP glitches (bold line) and radio pulsar glitches
(thin line) for (a) fractional frequency jump and (b) absolute frequency
jump (in Hz).  Radio pulsar glitch amplitudes are from an unpublished
catalog kindly supplied by A. Lyne.  AXP glitches included here are those
listed in Tables~\ref{table1708glitches} and \ref{table1841glitches}, the
2002 1E~2259+586 glitch \citep{kgw+03,wkt+04}, as well as a recent
unpublished 1E~2259+586 glitch (Dib et al., in prep.).
\label{fig:amp_dist}}
\end{figure}

Figure~\ref{fig:amp_dist} shows the fractional and non-fractional amplitude
distributions of radio pulsar and AXP glitches.  As is clear from the
figure, although the fractional glitch amplitudes of AXPs are generally
large by radio pulsar standards, the AXP absolute glitch amplitudes, more
directly related to the angular momentum transfer during the glitch, are
neither especially large nor especially small.  Thus, glitching in neutron
stars is clearly not correlated with frequency as some studies of radio
pulsars have suggested \citep{lsg00}.

Given the spectacular radiative outburst contemporaneous with the large 2002
1E~2259+586 glitch, we can speculate that larger angular momentum transfers
that occur in radio pulsars could result in even more dramatic outbursts in
affected AXPs, possibly like those seen in XTE~1810$-$197 \citep{ims+04} and
in the AXP candidate AX J1845$-$0258 \citep{tkgg06}.  Indeed a recent X-ray
burst observed from CXOU J164710.2$-$455216 \citep{mgc+07} has been claimed
to be accompanied by a very large ($\Delta \nu/\nu \simeq 6 \times 10^{-5}$,
$\Delta \nu \simeq 6 \times 10^{-6}$) glitch \citep{icd+07}, and AXP
1E~1048.1$-$5937 recently exhibited a large glitch and flux increase
\citep[][Dib et al., in prep.]{dkgw07}.  However the lack of any observed
radiative change in \eight\ around the time of its first observed glitch,
which was over a factor of two larger than that in 1E~2259+586 in terms of
absolute frequency jump, argues against this idea. Clearly, the data are
indicating that AXP glitches, even large ones, can be either radiatively
loud or quiet.

Glitch activity has been defined as
\begin{equation}
a_g = \frac{1}{\Delta t} \sum \frac{\Delta \nu}{\nu},
\label{eq:ag}
\end{equation}
where $\Delta t$ is the total observing span and the sum is over all
glitches, and includes decaying components \citep{ml90}. We refer to $a_g$
as fractional activity, since it involves the sum of fractional frequency
changes.  One can also define an absolute glitch activity,
\begin{equation}
A_g =\frac{1}{\Delta t} \sum \Delta \nu,
\label{eq:Ag}
\end{equation}
where the sum is over the absolute frequency changes
\citep[e.g.,][]{wmp+00}. The quantities $a_g$ and $A_g$, introduced for the
study of radio pulsars, are approximately interchangeable for those objects,
given that the range of frequencies encompassed by glitching radio pulsars
is relatively small.  By contrast, when considering AXPs and their much
smaller rotation frequencies, a comparison with radio pulsars for $a_g$ and
$A_g$ are very different \citep[see, e.g.,][]{hh99}. Also, for establishing
the average amount of spin-up imparted to the crust over time, the total
frequency increase at each epoch is relevant.  However, in some instances,
the quantity of interest is the {\it unrelaxed} portion of the glitch, i.e.,
the permanent frequency jump only.  In general, for radio pulsars, $Q$ is
small so this distinction is not important.  However for AXPs, given the
paucity of glitches we have observed thus far as well as the fact that
several, particularly the largest, of these have had large values of $Q$
(e.g., $Q\simeq 1$ for the second glitch seen in \seven), the distinction
between including the total frequency jump and only the unrelaxed frequency jump is
important.  We choose here to remain with convention and include the total
frequency jump at each epoch when calculating $a_g$ and $A_g$, although this
choice should be kept in mind.

With 8.7~yr of monitoring of \seven, we can now reasonably calculate glitch
activity parameters for this source using Equations~\ref{eq:ag} and
\ref{eq:Ag}.  If only counting the unambiguous glitches, $a_g = 2.9 \times
10^{-14}$~s$^{-1}$ and $A_g = 2.5 \times 10^{-15}$~s$^{-2}$. Including
candidate glitches only increases these numbers by $\sim$20\%. With three
glitches in 7.6~yr, \eight\ is evidently a very active glitcher as well.
Its glitch activity parameters are $a_g = 7.9 \times 10^{-14}$~s$^{-1}$ and
$A_g = 6.7 \times 10^{-15}$~s$^{-2}$. Indeed $A_g$ for \eight\ is the
highest glitch activity seen thus far in any neutron star, radio pulsar or
AXP, to our knowledge. We also calculated a tentative glitch activity for
AXP 1E~2259+586, for which we have observed two glitches, the well
documented one in 2002 \citep{kgw+03,wkt+04} and a second, smaller glitch
that occured very recently and had fractional amplitude $8.5\times 10^{-7}$
and no recovery (Dib et al., in prep.).  Using these events and given that
we have observed this source with {\it RXTE} for 9.4~yr, we find $a_g = 1.7
\times 10^{-14}$~s$^{-1}$ and $A_g = 2.4 \times 10^{-15}$~s$^{-1}$.

\begin{figure}
\includegraphics[scale=0.7]{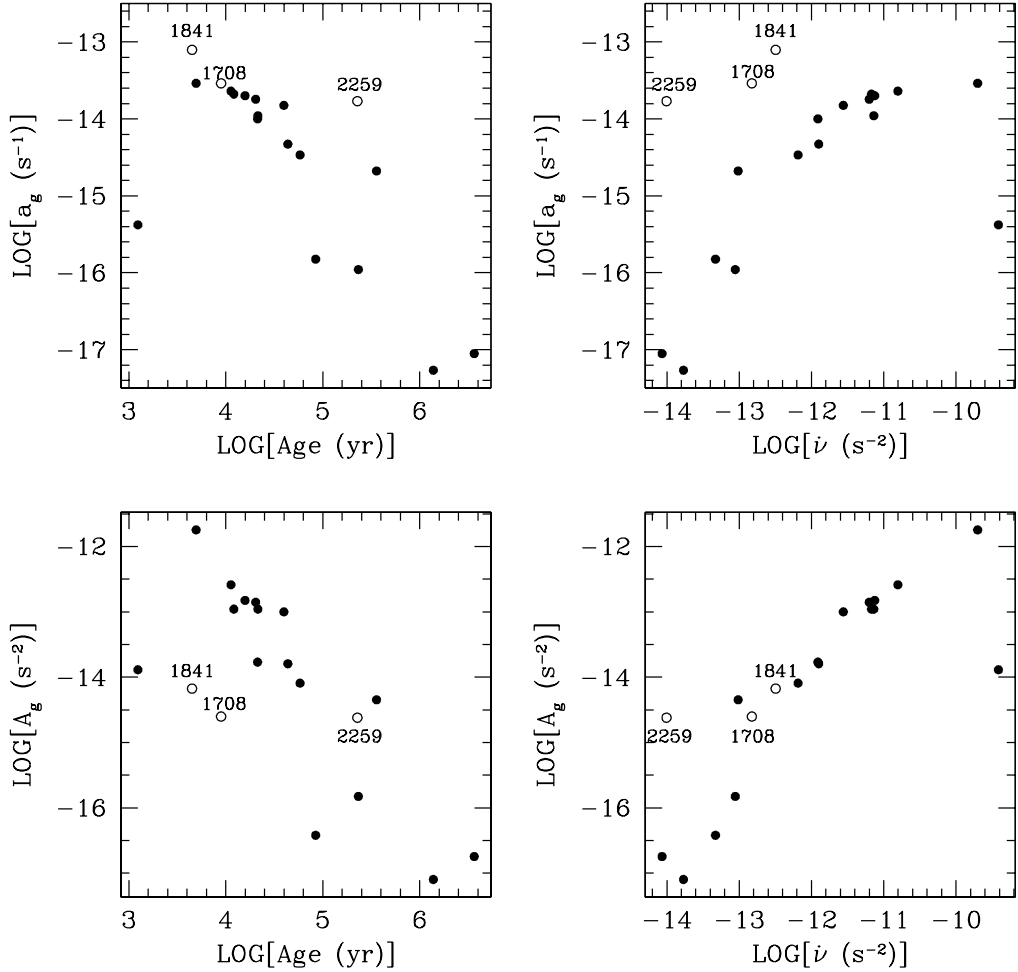}
\caption{
Activity parameters versus age (as estimated from $\nu/2\dot{\nu}$) and
versus $\dot{\nu}$ for radio pulsars and AXPs.  Fractional activity $a_g$ is
defined using the sum of the fractional frequency changes, while activity
$A_g$ is defined using the absolute frequency jumps. The only radio pulsars
included (filled circles) are those having exhibited three glitches or more
during continual (e.g., bi-monthly) monitoring, as recorded in the
unpublished glitch catalog kindly supplied by A. Lyne.  The AXPs included
here (open circles) are \seven, \eight, and 1E~2259+586. The latter has
glitched twice, once in 2002 \citep{kgw+03,wkt+04}, and once in 2007 (Dib et
al., in prep.).  Only unambiguous glitches were included for \seven; as the
candidate glitches are small, including them does not make a qualitative
difference.
\label{fig:actage}}
\end{figure}

We can plot these activities as a function of pulsar age (as estimated via
spin-down age $\nu/2\dot{\nu}$) and $\dot{\nu}$; see
Figure~\ref{fig:actage}. Previous authors have noted interesting
correlations on these plots for radio pulsars \citep[e.g.,][]{lsg00,wmp+00};
these are seen in our plots as well.  Note that upper limits for some radio
pulsars of relevant ages fall well below the apparent correlations
\citep[e.g.,][]{wmp+00}; we choose not to plot those because, as discussed
in that reference, a single glitch of average size would bring them roughly
in line with the correlation. Note that the radio pulsar outlier at small
age and high $\dot{\nu}$ in all plots is the Crab pulsar, long-known to
exhibit few and small glitches. We also looked for a trend in a plot of
$a_g$ or $A_g$ versus surface dipolar field (as estimated via $3.2\times
10^{19} \sqrt{P\dot{P}}$~G) but found none.

As a group, the AXPs do not especially distinguish themselves when either
activity, $a_g$ or $A_g$ is plotted versus spin-down age, though they do
increase the scatter.  This suggests a universal correlation with spin-down
age.  The same is true of $A_g$ plotted versus $\dot{\nu}$. However,
interestingly, the AXPs as a group all stand out on the diagram of $a_g$
versus $\dot{\nu}$ (Fig.~\ref{fig:actage}), such that for similar spin-down
rates, their fractional glitch activities are much larger than in radio
pulsars.

\begin{figure}
\includegraphics[scale=0.6]{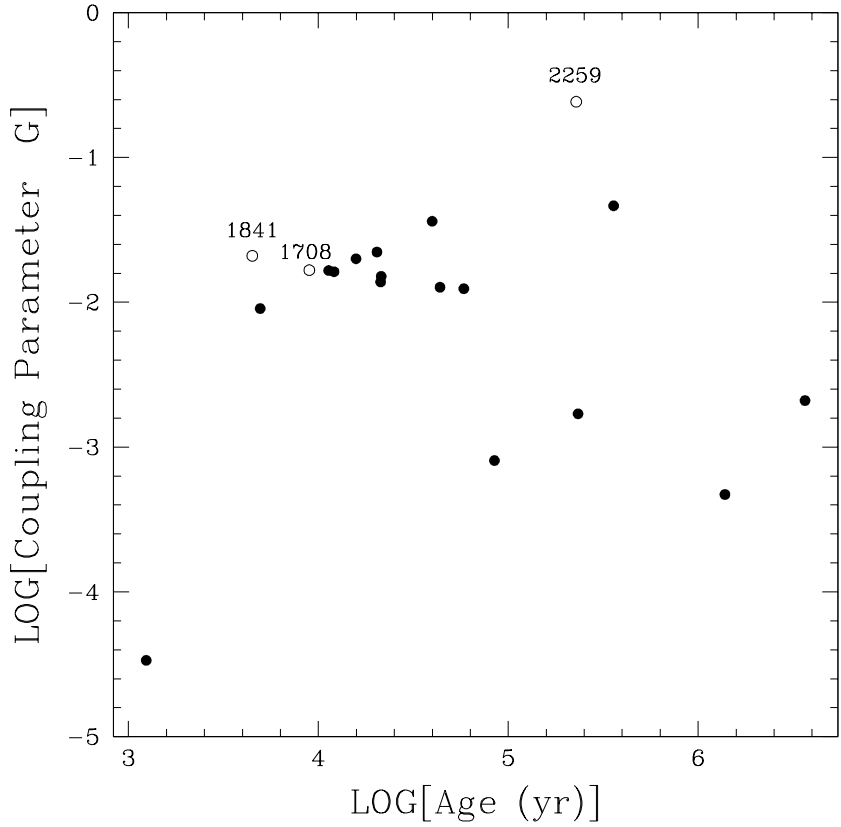}
\caption{
``Coupling parameter'' $G$ (as defined by \protect\citep{lel99}; see
\S\ref{sec:disc_glitch}) as a function of spin-down age ($\nu/2\dot{\nu}$).
Solid points are radio pulsars with three or more observed glitches, as
recorded in the unpublished catalog of A. Lyne.  Open circles are AXPs
1E~2259+586, \seven, and \eight.
\label{fig:gage}}
\end{figure}

\citet{lel99} argued that $a_g$ provides a strict lower limit on the
fraction of the moment of inertia of the neutron star that resides in the
angular momentum reservoir (generally assumed to be the crustal superfluid)
tapped during spin-up glitches, $I_{res}$. They showed that $I_{res}/I_c
\geq \nu a_g/|{\dot{\nu}}| \equiv G$, where $I_c$ is the moment of inertia
of the crust and all components strongly coupled to it, and $G$ is a
``coupling parameter.'' For radio pulsars, they argued for a universal $G$
that implies $I_{res}/I_c \geq 0.014$. It is interesting to ask whether this
same apparently universal relationship holds for AXPs.
Figure~\ref{fig:gage} shows $G$ plotted versus age for radio pulsars and
AXPs.  As is clear, the \citet{lel99} relation seems to hold for the radio
pulsars, even with increased glitch statistics.  Also, AXPs \seven\ and
\eight\ lie among the radio pulsars, suggesting similar reservoir fractions.
However the outlier point, 1E~2259+586, has $G=0.25$, much larger than the
others. Admittedly, for this AXP, $a_g$ is estimated from two glitches only,
with the 2002 event greatly dominating, so the values are tentative.  Still,
the large $G$, if real, suggests that at least $\sim$25\% of the stellar
moment of inertia is in the angular momentum reservoir \citep[see
also][]{wkt+04}.  We note that the analysis of \citet{lel99} ignores
recovery, important for the 2002 1E~2259+586 glitch, which dominates its
$a_g$.  However in the 2002 glitch, the recovery fraction was only
$\sim$19\%, so even accounting for recovery, $G$ for 1E~2259+586 is
surprisingly high.

As described by \citet{kgw+03} and \citet{wkt+04}, the 2002 1E~2259+586
glitch was unusual when compared with those of radio pulsars. Specifically
the combination of the recovery time scale and the large recovery fraction
$Q$ conspired to make the pulsar spin down, for over two weeks post-glitch,
at over twice its long-term average spin-down rate.  Although spin-down rate
enhancements post-glitch are often seen in radio pulsars
\citep[e.g.,][]{fla90}, they usually amount to only a few percent.  A
remarkably large post-glitch spin-down rate enhancement was seen also in the
second glitch of \seven\ and the first observed glitch of \eight, though to
a lesser degree than in 1E~2259+586. Of course a much larger increase in
spin-down rate post-glitch in \seven\ or \eight\ could have been missed due
to our sparse sampling.

One way to quantify the enhanced spin-down more precisely is using
Equation~\ref{eq:glitch} at $t=0$, and noticing that the instantaneous
spin-down rate at the glitch epoch due to the exponential recovery is given
by $\Delta\nu_d / \tau$.  Comparing this quantity for the AXP glitches that
show recovery with the pre-glitch time-averaged spin-down rate $\dot{\nu}$,
we find that for 1E~2259+586 $\Delta\nu_d / \tau = (8.2 \pm 0.6) \dot{\nu}$,
$\Delta\nu_d / \tau = (0.64 \pm 0.6) \dot{\nu}$ for \seven, and $\Delta\nu_d
/ \tau = (0.75 \pm 0.08) \dot{\nu}$ for \eight, all very large by radio
pulsar standards.

The increase in spin-down rate post-glitch, at least for radio pulsars, is
generally attributed to a decoupling of a small percentage of the moment of
inertia of the star, usually presumed to be part of the crustal superfluid
\citep[e.g.,][]{pa85}, with constant external torque.  If the observed AXP
recoveries and temporarily enhanced spin-down rates were interpreted in the
same way, it would imply that very large fractions (ranging from 0.4 to 0.9)
of the moment of inertia of the star decoupled at the glitch, much larger
than the crustal superfluid is reasonably expected to comprise, for any
interior equation of state.  To avoid this problem, \citet{wkt+04} suggested
that a pre-glitch rotational lag between the crust and superfluid might have
temporarily reversed at the glitch \citep[see also][]{apc90}. Then the
observed larger spin-down rate post-glitch would be due to the crust
transferring angular momentum back to the superfluid in order to reestablish
equilibrium.

In glitches, the equilibrium angular velocity lag between the crust and more
rapidly rotating crustal superfluid is thought to be the origin of glitches.
This lag is proposed to develop because the crustal superfluid's angular
momentum vortices, in many models \citep[e.g.,][]{aaps84b}, become pinned to
crustal nuclei and hence are hindered from moving outward as the star's
crust and associated components are slowed by the external torque. How this
lag could reverse is puzzling.  \citet{wkt+04} suggested that a twist of
magnitude $10^{-2}$~rad of a circular patch of crust offset in azimuth from
the rotation axis could result in sufficient spin-down of the crustal
superfluid to account for the properties of the 2002 glitch in 1E~2259+586.
They noted further that such a twist also produces X-rays of the luminosity
observed in that outburst \citep{tlk02}. The absence of any significant
radiative changes at the time of largest glitches in \seven\ and \eight\ is
thus problematic for the crustal twist, and hence lag reversal, model.  We
note that it has been argued independently that a similar suggested lag
reversal between crust and crustal superfluid in the Vela pulsar is
unphysical \citep{jah05}.

We also note that the large and long-term increase in the magnitude of
$\dot{\nu}$ following the large glitch in \eight\ (see
\S\ref{sec:timing1841}) is also interesting.  \citet{accp93} showed that,
ignoring transient terms, $I_{res}/I_c \geq \Delta\dot{\nu}/\dot{\nu}$.  For
the large \eight\ glitch, this implies $I_{res}/I_c \geq 0.1$, much larger
than has been seen in any radio pulsar.

One possibility that can explain the large $G$ for 1E~2259+586, the large
transient increases in the magnitude of $\dot{\nu}$ in all three large
glitches, as well as the large extended $\dot{\nu}$ change in the first
\eight\ glitch, is that core superfluid is somehow involved, as it is expected to
carry the bulk of the moment of inertia.  We note that core glitches have
been discussed in the radio pulsar context for some time, albeit for very
different reasons.  Although \citet{als84} argued that the crust and core
should be strongly coupled on very short time scales, \citet{jon98} found
that crustal pinning of superfluid vortices cannot be occuring in neutron
stars, because the maximum pinning force is orders of magnitude smaller than
the estimated vortex Magnus force.  \citet{dp03c} argue that crustal vortex
pinning cannot occur for independent reasons. If these authors are correct,
pulsar glitches would generally not originate in the crust.

Why would the clearest evidence for core glitches come from AXPs? The
interaction between vortices and quantized magnetic flux tubes in a core
superfluid could provide resistance to outward motion of vortices
\citep{jon98,rzc98,jon02}.  The interior magnetic field playing a role in
vortex pinning as studied by \citet{rzc98} could help explain why such
unusual glitch recoveries are seen preferentially in AXPs, which appear to
have much larger magnetic fields than conventional radio pulsars.  Perhaps
the larger field, which implies a higher density of flux tubes, can
effectively pin more superfluid vortex lines in a magnetar core; with
greater magnetic activity, sudden magnetic reconfigurations would result in
large core vortex reconfigurations. We note that \citet{klc00} and
\citet{dis+03} argued that the \citet{rzc98} model must be inapplicable to
AXPs as that model predicts no glitches for periods greater than $\sim
0.7$~s.  However a more careful reading of \citet{rzc98} reveals that this
prediction does not apply for magnetar-strength magnetic fields.

As pointed out by \citet{klc00}, the high temperatures of AXPs, as measured
from their X-ray spectra, are at odds with the glitch observations.  This is
because in the crustal pinning models, the pinning force is highly
temperature dependent, such that vortex lines can creep outward much more
easily when the temperature is high \citep[e.g.,][]{acp89}.  This has long
been the explanation \citep[e.g.,][]{ai75} for the difference between the
Crab and Vela pulsar glitch behaviors: the hotter Crab pulsar glitches less
frequently and with smaller frequency jumps because its vortex array can
move outward more smoothly.  If this were true, AXPs, having measured
effective temperatures much higher than the Crab pulsar, should glitch less
frequently and with smaller glitch amplitudes than the Crab pulsar, clearly
not what is observed.  If we abandon the crustal glitch model (at least in
AXP glitches) the absence of the expected temperature dependence and the
observed universal age correlation could be explained. For example, as
discussed by \citet{jon98}, perhaps relatively smooth outward motion of core
vortices could take place before the magnetic flux distribution necessary
for impeding them has developed.  In this picture, magnetic field
distribution, not temperature, is the age-associated property that is a
primary factor determining glitch behavior.

Finally, \citet{cri+07} argued, on the basis of seven observations of
\seven\ obtained over $\sim$10~yr, that observed spectral and flux
variations were correlated with glitch epoch.  They predict, given apparent
flux increases seen in mid-2004 and mid-2005, that a glitch should occur
soon thereafter (see Fig.~\ref{figure1708flux}).  Indeed as we have shown
(see Table~\ref{table1708glitches}), an unambiguous glitch occured just
following their mid-2005 observation. On the other hand, the first candidate
glitch (Table~\ref{table1708cands}) occured when the total flux was very low
and apparently declining.  Thus, if there is a causal connection between
long-term flux variations in \seven\ and glitches, either this candidate
glitch is not a true glitch, or accompanying radiative changes are only
relevant to large glitches. The sparsity of the total flux measurements,
along with the relatively short time scale of the pulsed flux variations we
report in \seven, suggest that more dense total flux monitoring could reveal
yet unseen fluctuations that are not glitch-associated. The flux variability
of \seven\ is discussed further below.

\subsection{Radiative Changes}
\label{sec:disc_rad}

The approximate stabilities of the pulsed fluxes of \seven\ and \eight\ (see
\S\ref{sec:flux}) are in contrast to those seen for AXPs 1E~2259+586 and
1E~1048.1$-$5937, both of which have shown large pulsed flux variations
\citep{kgw+03,wkt+04,gk04}, and even 4U~0142+61 which has shown a slow
pulsed flux increase with time \citep{dkg07}. It is interesting that the
phase-averaged flux of \seven\ has been reported to be highly variable
(\citealp{roz+05,cri+07} and Fig.~\ref{figure1708flux}), with changes as
large as $\sim$60\% in 2004-2005, while the pulsed flux is not, with maximum
contemporaneous change of $<15$\%.  Note that this conclusion holds even
when the phase-averaged fluxes, reported in the 0.5--10~keV band, are
converted to 2--10~keV, that used for our {\it RXTE} observations.  This
suggests an anti-correlation between pulsed fraction and total flux that
acts to ensure that the pulsed flux is roughly constant. If so, pulsed flux
is not a good indicator of total energy output for \seven\.  A similar
anti-correlation between total flux and pulsed fraction has been reported
for AXP 1E~1048$-$5937 \citep{tmt+05,gkw06,gkwd07}, although in that case
the pulsed flux does not remain constant \citep{gk04}, but follows the
phase-averaged flux, just with lower dynamic range (Tam et al., submitted).
For \seven, the exactness of the anti-correlation is perhaps surprising.  It
could be that all the phase-averaged flux variations are in the 0.5--2~keV
band, invisible to {\it RXTE}.  However this would not jibe with the
reported correlation of the phase-averaged fluxes with power-law index
\citep{roz+05,cri+07}.  We also note that in the phase-averaged flux
analysis, the equivalent hydrogen column $N_H$ was allowed to vary from
observation to observation, rather than being held fixed at a constant. This
inconsistency could bias the comparison, though likely not by a large
amount.  It is tempting to question the relative calibrations of the
different instruments used to measure the phase-averaged flux of \seven, as
the greatest dynamic range is implied by lone {\it XMM-Newton} and {\it
Swift} observations, and even the two {\it Chandra X-ray Observatory}
observations were obtained with different instruments.  Still, admittedly,
relative systematic calibration uncertainties are not expected to yield a
$>$50\% dynamic range, as is reported.  Regular monitoring with a single
imaging instrument could settle this issue.

Changes in pulse profile seem to be generic in AXPs and, as discussed
above, are not always correlated with glitches, although in some cases,
e.g., 1E~2259+586, 4U~0142+61, and possibly \seven, they are.  Spectrally
no clear pattern has emerged.  In 1E~2259+586, the pulse profile changes
following its 2002 event were broadband \citep{wkt+04}, while in 4U~0142+61
\citep{dkg07} and \eight, they appear more prominent at soft energies.  In
\seven, the changes are more apparent in the hard band. In the context of
the magnetar model, this hints at crustal motions and surface activity,
possibly coupled with magnetospheric activity, with the exact observational
manifestation dependent on a variety of factors ranging from viewing
geometry to magnetospheric scattering optical depth.  Whether ultimately
this specific phenomenon will provide insights into the physics of magnetars
remains to be seen.

\section{Summary}
\label{sec:sum}

We have reported on long-term {\it RXTE} monitoring of AXPs \seven\ and
\eight, which has allowed us to study these sources' timing, pulsed flux,
and pulse profile evolutions.

We have discovered four new AXP spin-up glitches, one in \seven\ and three
in \eight, plus three new glitch candidates in \seven. Nearly all of the
``classical'' AXPs have now been seen to glitch, clearly demonstrating that
this behavior is generic to the class.  Moreover, in terms of fractional
frequency increases, AXPs are among the most actively glitching neutron
stars known.  Further, unlike radio pulsar glitches, AXP glitches appear to
come in two varieties: those that, like radio pulsars, are radiatively quiet
in pulsed flux, and those that are, unlike radio pulsars, radiatively loud,
including correlated sudden flux increases and pulse profile changes.  Thus
far there is no clear correlation between AXP glitch size and whether or not
it will be radiatively loud or quiet -- two of the largest AXP glitches thus
far were quiet. We have found a substantial long-term increase in the
magnitude of the spin-down rate in the largest glitch from \eight, and have
also shown that large AXP glitches often have recoveries that are unusual
compared with those seen in radio pulsars. Specifically, their spin-down
rates in the days and sometimes weeks after a glitch are
significantly larger in absolute value than their long-term spin-down rate.
This latter effect may indicate a temporary reversal in the crust/crustal
superfluid lag at the time of the glitch, or possibly more plausibly, and
certainly more intriguingly, glitches of the core, which could explain the
transient and extended $\dot{\nu}$ increases, as well as the large $G$ value
for 1E~2259+586.

Radiatively, we have found that the pulsed fluxes of \seven\ and \eight\ are
both fairly steady with time.  This is perhaps surprising in light of the
large changes in phase-averaged flux that have been reported for \seven, and
suggests, unless the latter are affected by systematic calibration
uncertainties, that pulsed flux for this source, as for AXP 1E~1048.1$-$5937
\citep{tmt+05}, is not a good indicator of AXP X-ray output.  Also, the
pulse profiles of \seven\ and \eight\ both evolve; such evolution appears
also to be a generic property of AXPs. However, no clear patterns in AXP
pulse profile changes have yet emerged beyond occasional correlation with
glitches.  Hopefully further monitoring will shed physical light on this
phenomenon.

\acknowledgments

We are grateful to Andrew Lyne for providing his unpublished glitch catalog.
We thank Andrew Cumming, Chris Thompson and Pete Woods for useful comments.
This work was supported by the Natural Sciences and Engineering Research
Council (NSERC) PGSD scholarship to RD. FPG is supported by the NASA
Postdoctoral Program administered by Oak Ridge Associated Universities at
NASA Goddard Space Flight Center. Additional support was provided by NSERC
Discovery Grant Rgpin 228738-03, NSERC Steacie Supplement Smfsu 268264-03,
FQRNT, cifar, and CFI.  VMK holds the Lorne Trottier in Astrophysics and
Cosmology and a Canada Research Chair in Observational Astrophysics.


\clearpage

\clearpage

\clearpage

\clearpage

\clearpage

\clearpage

\clearpage

\clearpage

\clearpage

\clearpage

\clearpage

\clearpage

\clearpage

\clearpage

\clearpage

\clearpage

\clearpage
\begin{deluxetable}{ccccccc}
\tabletypesize{\footnotesize}
\tablewidth{469.0pt}
\tablecaption
{Summary of {\em{RXTE}} Observations of \seven\
\label{tableobs1}}
\tablehead
{
\colhead{Obs.} &
\colhead{Typical} &
\colhead{Typical} &
\colhead{No. of} &
\colhead{Total} &
\colhead{First -- Last}\\
\colhead{Cycle} &
\colhead{Exp.\tablenotemark{a}} &
\colhead{Separation\tablenotemark{a}} &
\colhead{Obs.\tablenotemark{b}} &
\colhead{Exp.\tablenotemark{c}} &
\colhead{MJD\tablenotemark{d}} &
\colhead{First Date $-$ Last Date} \\
&
\colhead{(ks)} &
\colhead{(days)} &
&
\colhead{(ks)} &
}
\startdata
3  & 2.5 & 15  & 29 & 75  & 50825.7$-$51186.7 & 01/12/1998$-$01/08/1999\\
4  & 3   & 24  & 20 & 60  & 51215.7$-$51614.1 & 02/06/1999$-$03/11/2000\\
5  & 3   & 24  & 13 & 40  & 51655.7$-$52041.5 & 04/21/2000$-$05/12/2001\\
6  & 3   & 29  & 13 & 40  & 52049.5$-$52325.6 & 05/20/2001$-$02/20/2002\\
7  & 5.5 & 23  & 12 & 65  & 52366.5$-$52718.7 & 04/02/2002$-$03/20/2003\\
8  & 1.8 & 5   & 58 & 105 & 52745.7$-$53058.6 & 04/16/2003$-$02/23/2004\\
9  & 2   & 5   & 70 & 135 & 53063.1$-$53429.1 & 02/28/2004$-$02/28/2005\\
10 & 2   & 8   & 47 & 90  & 53435.1$-$53791.6 & 03/06/2005$-$02/25/2006\\
11\tablenotemark{e} & 2   & 7   & 32 & 60  & 53799.0$-$54015.4 & 03/05/2006$-$10/07/2006\\
\enddata
\tablenotetext{a}{The exposure and separation are approximate.  Note that the PCA effective area changed with time primarily due the reduction of the average number of PCUs operational during an integration.  This effect is not incorporated in the tabulated integration times.}
\tablenotetext{b}{When the last digits of the observation ID of two successive data sets
are different, the two data sets are considered separate observations.}
\tablenotetext{c}{The total exposure does not include Earth occultation periods.}
\tablenotetext{d}{First MJD and Last MJD are the epochs, in Modified Julian Days, 
of the first and the last observations in a Cycle.}
\tablenotetext{e}{Cycle 11 not yet completed.}
\end{deluxetable}

\clearpage
\begin{deluxetable}{ccccccc}
\tabletypesize{\footnotesize}
\tablewidth{499.0pt}
\tablecaption
{
Summary of  {\em{RXTE}} Observations of \eight\ 
\label{tableobs2}
}
\tablehead
{
\colhead{Obs.} &
\colhead{Typical} &
\colhead{Typical} &
\colhead{No. of} &
\colhead{Total} &
\colhead{First -- Last}\\
\colhead{Cycle} &
\colhead{Exp.\tablenotemark{a}} &
\colhead{Separation\tablenotemark{a}} &
\colhead{Obs.\tablenotemark{b}} &
\colhead{Exp.\tablenotemark{c}} &
\colhead{MJD\tablenotemark{d}} &
\colhead{First Date $-$ Last Date} \\
&
\colhead{(ks)} &
\colhead{(days)} &
&
\colhead{(ks)} &
}
\startdata
4  & 4.5 & 27 & 26 & 120 & 51224.4$-$51597.3 & 02/15/1999$-$02/23/2000\\
5  & 4.5 & 38 & 16 & 70  & 51644.7$-$51976.9 & 04/10/2000$-$03/08/2001\\
6  & 7   & 27 &  8 & 50  & 52001.6$-$52300.1 & 04/02/2001$-$01/26/2002\\
7  & 12  & 45 &  7 & 80  & 52349.8$-$52666.0 & 03/16/2002$-$01/27/2003\\
8  & 4.5 & 20 & 17 & 80  & 52726.8$-$53052.9 & 03/28/2003$-$02/17/2004\\
9  & 4.5 & 20 & 19 & 80  & 53073.7$-$53413.3 & 03/09/2004$-$02/12/2005\\
10 & 5   & 14 & 31 & 130 & 53440.0$-$54153.9 & 03/11/2005$-$02/22/2007\\
11\tablenotemark{e} & 5 & 14 & 12 & 60 & 53800.9$-$53970.6 & 03/06/2006$-$08/23/2006\\
\enddata
\tablenotetext{a}{The exposure and separation are approximate.  Note that the PCA effective area changed with time primarily due the reduction of the average number of PCUs operational during an integration.  This effect is not incorporated in the tabulated integration times.}
\tablenotetext{b}{When the last digits of the observation ID of two successive data sets
are different, the two data sets are considered separate observations.}
\tablenotetext{c}{The total exposure does not include Earth occultation periods.}
\tablenotetext{d}{First MJD and Last MJD are the epochs, in Modified Julian Days,
of the first and the last observations in a Cycle.}
\tablenotetext{e}{Cycle 11 not yet completed.}
\end{deluxetable}

\clearpage
\begin{deluxetable}{lccccccc}
\tabletypesize{\tiny}
\tablewidth{510.0pt}
\tablecaption
{
Long-Term Spin Parameters for RXS~J170849.0$-$400910\tablenotemark{a}
\label{table1708big}
}
\tablehead
{
&
\colhead{Ephemeris A} &
\colhead{Ephemeris B} &
\colhead{Ephemeris C} &
\colhead{Ephemeris D} &
\colhead{Ephemeris E} &
\colhead{Ephemeris F} &
\colhead{Ephemeris G} \\
\colhead{Parameter} &
\colhead{Spanning MJD} &
\colhead{Spanning MJD} &
\colhead{Spanning MJD} &
\colhead{Spanning MJD} &
\colhead{Spanning MJD} &
\colhead{Spanning MJD} &
\colhead{Spanning MJD} \\
&
\colhead{50826$-$51418} &
\colhead{51446$-$51996} &
\colhead{52036$-$52960} &
\colhead{53010$-$53325} &
\colhead{53377$-$53548} &
\colhead{53556$-$53631} &
\colhead{53638$-$54015}
}
\startdata
MJD start & 50826.078 & 51446.610 & 52035.655 & 53010.094 & 53377.133 & 53555.734 & 53638.033 \\
MJD end & 51418.374 & 51995.680 & 52960.186 & 53325.061 & 53547.811 & 53631.161 & 54015.487 \\
TOAs & 39 & 19 & 74 & 69 & 29 & 13 & 49 \\
$\nu$ (Hz) & 0.090913818(2) & 0.090906071(3) & 0.090906089(3) & 0.090892731(13) & 0.090887608(18)  & 0.090885281(8) & 0.090884082(9) \\
$\dot{\nu}$ (10$^{-13}$ Hz s$^{-1}$) & $-$1.583(3) & $-$1.574(2) & $-$1.565(6) & $-$1.40(5) & $-$1.19(8) & $-$1.70(2) & $-$1.58(3) \\
$\ddot{\nu}$ (10$^{-22}$ Hz s$^{-2}$) & $-$1.4(3) & 0.36(9) & $-$8.9(1.8) & $-$44(11) & $-$131(23) & $-$ & $-$8(7) \\
$d^{3}\nu/dt^{3}$ (10$^{-28}$ Hz s$^{-3}$) & $-$0.056(9) & $-$ & 1.5(3) & 5.5(1.6) & 15(3) & $-$ & 1.4(9)\\
$d^{4}\nu/dt^{4}$ (10$^{-35}$ Hz s$^{-4}$) & $-$ & $-$ & $-$1.4(3) & $-$3.1(1.0) & $-$ & $-$& $-$0.8(6) \\
$d^{5}\nu/dt^{5}$ (10$^{-43}$ Hz s$^{-5}$) & $-$ & $-$ & 7.4(1.5) & $-$ & $-$ & $-$ & $-$ \\
$d^{6}\nu/dt^{6}$ (10$^{-50}$ Hz s$^{-6}$) & $-$ & $-$ & $-$1.8(4) & $-$ & $-$ & $-$ & $-$ \\
$\Delta \nu _{d}$\tablenotemark{b} (Hz) & $-$ & $-$ & 36(3)$\times$10$^{-08}$ & $-$ & $-$ & $-$ & $-$ \\
$t_d$\tablenotemark{b} (days) & $-$ & $-$ & 43(2) & $-$ & $-$ & $-$ & $-$ \\
Epoch (MJD) & 51445.3846 & 52016.48413 & 52016.48413 & 52989.8475 & 53366.3150 & 53549.15095 & 53635.6772 \\
RMS residual (phase) & 0.0079 & 0.0150 & 0.0154 & 0.0132 & 0.0142 & 0.0112 & 0.0154 \\
\enddata
\tablenotetext{a}{Numbers in parentheses are TEMPO-reported 1$\sigma$ uncertainties.}
\tablenotetext{b}{Parameters held fixed at values determined from local glitch fits as
shown in Table~\ref{table1708glitches}.}
\end{deluxetable}

\clearpage
\begin{deluxetable}{lccc}
\tabletypesize{\small}
\tablewidth{420.0pt}
\tablecaption
{
Local Ephemerides of RXS~J170849.0$-$400910 Near Glitch Epochs\tablenotemark{a}
\label{table1708glitches}
}
\tablehead
{
&
\colhead{Ephemeris} &
\colhead{Ephemeris} &
\colhead{Ephemeris} \\
\colhead{Parameter} &
\colhead{Near Glitch~1} &
\colhead{Near Glitch~2} &
\colhead{Near Glitch~3} \\
}
\startdata
MJD range & 51186.503$-$51614.187 & 51614.185$-$52366.663  & 53465.392$-$53631.161 \\
TOAs & 22 & 29 & 26 \\
Epoch (MJD) & 51445.3846 & 52016.48413 & 53549.15095 \\
$\nu$ (Hz) & 0.090913822(2) & 0.090906068(2) & 0.090885035(9) \\
$\dot{\nu}$ (10$^{-13}$ Hz s$^{-1}$) & $-$1.5714(14) & $-$1.5797(11) & $-$1.67(2) \\
Glitch Epoch (MJD) & 51445.3846 & 52016.48413 & 53549.15095 \\
$\Delta\nu$ (Hz) & 5.1(3)$\times$10$^{-8}$ & 2.2(4)$\times$10$^{-8}$ & 24.6(9)$\times$10$^{-8}$ \\
$\Delta\dot{\nu}$ (Hz s$^{-1}$) & $-$0.8(4)$\times$10$^{-15}$ & $-$1.1(2)$\times$10$^{-15}$ & $-$2(2)$\times$10$^{-15}$ \\
$\Delta \nu _{d}$ (Hz) & $-$ & 36(3)$\times$10$^{-8}$ & $-$ \\
$t_d$ (days) & $-$ & 43(2) & $-$ \\
RMS residual (phase) & 0.0102 & 0.0193 & 0.0140 \\
\enddata
\tablenotetext{a} {Numbers in parentheses are TEMPO-reported 1$\sigma$
uncertainties. }
\end{deluxetable}

\clearpage
\begin{deluxetable}{lccc}
\tabletypesize{\small}
\tablewidth{420.0pt}
\tablecaption
{
Local Ephemerides of RXS~J170849.0$-$400910 Near Candidate Glitch Epochs\tablenotemark{a}
\label{table1708cands}
}
\tablehead
{
&
\colhead{Ephemeris} &
\colhead{Ephemeris} &
\colhead{Ephemeris} \\
\colhead{Parameter} &
\colhead{Near Candidate~1} &
\colhead{Near Candidate~2} &
\colhead{Near Candidate~3} \\
}
\startdata
MJD range & 52745.790$-$53140.604 & 53229.271$-$53456.688  & 53562.209$-$53785.652 \\
TOAs & 78 & 38 & 28 \\
Epoch (MJD) & 529898475 & 53366.3150 & 53635.6772 \\
$\nu$ (Hz) & 0.0908927493(18) & 0.090887617(5) & 0.090884020(8) \\
$\dot{\nu}$ (10$^{-13}$ Hz s$^{-1}$) & $-$1.5842(15) & $-$1.570(7) & $-$1.67(2) \\
Glitch Epoch (MJD) & 529898475 & 53366.3150 & 53635.6772 \\
$\Delta\nu$ (Hz) & 2.8(4)$\times$10$^{-8}$ & 5.2(6)$\times$10$^{-8}$ & 6.7(3)$\times$10$^{-8}$ \\
$\Delta\dot{\nu}$ (Hz s$^{-1}$) & 0\tablenotemark{b} & $-$1.9(1.3)$\times$10$^{-15}$ & 6.0(5)$\times$10$^{-15}$ \\
$\Delta \nu _{d}$ (Hz) & $-$ & $-$ & $-$ \\
$t_d$ (days) & $-$ & $-$ & $-$ \\
RMS residual (phase) & 0.0153 & 0.0099 & 0.0110 \\
\enddata
\tablenotetext{a} {Numbers in parentheses are TEMPO-reported 1$\sigma$
uncertainties. }
\tablenotetext{b} {Entries with the value `0' are consistent with being zero.}
\end{deluxetable}

\clearpage
\begin{deluxetable}{lccc}
\tabletypesize{\small}
\tablewidth{420.0pt}
\tablecaption
{
Alternate Ephemerides of RXS~J170849.0$-$400910 Near Candidate Glitch Epochs\tablenotemark{a}
\label{table1708alt}
}
\tablehead
{
&
\colhead{Ephemeris} &
\colhead{Ephemeris} &
\colhead{Ephemeris} \\
\colhead{Parameter} &
\colhead{Near Candidate~1} &
\colhead{Near Candidate~2} &
\colhead{Near Candidate~3} \\
}
\startdata
MJD range & 52745.790$-$53140.604 & 53229.271$-$53456.688  & 53562.209$-$53785.652 \\
TOAs & 75 & 38 & 28 \\
Epoch (MJD) & 52989.8475 & 53366.3150 & 53635.6772 \\
$\nu$ (Hz) & 0.0908928173(6) & 0.0908876402(13) & 0.090884059(2) \\
$\dot{\nu}$ (10$^{-13}$ Hz s$^{-1}$) & $-$1.556(3) & $-$1.492(6) & $-$1.485(16) \\
$\ddot{\nu}$ (10$^{-22}$ Hz s$^{-2}$) & 1.2(4) & 4(2) & 0\tablenotemark{b} \\
$d^{3}\nu/dt^{3}$ (10$^{-28}$ Hz s$^{-3}$) & $-$0.66(16) & $-$6.6(1.4) & $-$19(4) \\
$d^{4}\nu/dt^{4}$ (10$^{-35}$ Hz s$^{-4}$) & $-$1.0(4) & $-$18(6) & 52(12) \\
RMS residual (phase) & 0.0146 & 0.0106 & 0.0152 \\
\enddata
\tablenotetext{a} {Numbers in parentheses are TEMPO-reported 1$\sigma$
uncertainties. }
\tablenotetext{b} {Entries with the value `0' are consistent with being zero.}
\end{deluxetable}

\clearpage
\begin{deluxetable}{lccccc}
\tabletypesize{\tiny}
\tablewidth{410.0pt}
\tablecaption
{
Long-Term Spin Parameters for 1E~1841$-$045\tablenotemark{a}
\label{table1841big}
}
\tablehead
{
&
\colhead{Ephemeris A} &
\colhead{Ephemeris B} &
\colhead{Ephemeris B2} &
\colhead{Ephemeris C} &
\colhead{Ephemeris D} \\
\colhead{Parameter} &
\colhead{Spanning MJD} &
\colhead{Spanning MJD} &
\colhead{Spanning MJD} &
\colhead{Spanning MJD} &
\colhead{Spanning MJD} \\
&
\colhead{51225$-$52438} &
\colhead{52460$-$52981} &
\colhead{52610$-$52981} &
\colhead{53030$-$53816} &
\colhead{53829$-$53983}
}
\startdata
MJD start & 51224.538 & 52460.000 & 52610.313 & 53030.093 & 53828.808 \\
MJD end & 52437.712 & 52981.186 & 52981.186 & 53815.842 & 53983.431 \\
TOAs & 53 & 19 & 17 & 54 & 11 \\
$\nu$ (Hz) & 0.0849253002(9) & 0.084904428(7) & 0.084889922(3) & 0.084890135(6) & 0.084868767(7) \\
$\dot{\nu}$ (10$^{-13}$ Hz s$^{-1}$) & $-$2.9940(10) & $-$3.176(2) & $-$3.179(2) & $-$3.354(7) & $-$2.833(9) \\
$\ddot{\nu}$ (10$^{-22}$ Hz s$^{-2}$) & 3.30(14) & $-$ & $-$ & 16.4(4) & $-$ \\
$d^{3}\nu/dt^{3}$ (10$^{-29}$ Hz s$^{-3}$) & 0.9(2) & $-$ & $-$ & $-$2.81(11) & $-$ \\
$d^{4}\nu/dt^{4}$ (10$^{-36}$ Hz s$^{-4}$) & $-$2.3(2) & $-$ & $-$ & $-$ & $-$ \\
$d^{5}\nu/dt^{5}$ (10$^{-43}$ Hz s$^{-5}$) & 1.5(3) & $-$ & $-$ & $-$ & $-$ \\
$d^{6}\nu/dt^{6}$ (10$^{-51}$ Hz s$^{-6}$) & $-$3(2) & $-$ & $-$ & $-$ & $-$ \\
$\Delta \nu _{d}$\tablenotemark{b} (Hz) & $-$ & 8.1(6)$\times$10$^{-7}$ & $-$ & $-$ & $-$  \\
$t_d$\tablenotemark{b} (days) & $-$ & 43(3) & $-$ & $-$ & $-$  \\
Epoch (MJD) & 51618.000 & 52464.00448 & 52997.0492 & 52997.0492 & 53823.9694 \\
RMS residual (phase) & 0.029 & 0.025 & 0.028 & 0.033 & 0.022 \\
\enddata
\tablenotetext{a} {Numbers in parentheses are TEMPO-reported 1$\sigma$
uncertainties. } 
\tablenotetext{b}{Parameters held fixed at values determined from local glitch fits as
shown in Table~\ref{table1841glitches}.}
\end{deluxetable}

\clearpage
\begin{deluxetable}{lccc}
\tabletypesize{\small}
\tablewidth{420.0pt}
\tablecaption
{
Local Ephemerides of 1E~1841$-$045 Near Glitch Epochs\tablenotemark{a}
\label{table1841glitches}
}
\tablehead
{
&
\colhead{Ephemeris} &
\colhead{Ephemeris} &
\colhead{Ephemeris} \\
\colhead{Parameter} &
\colhead{Near Glitch~1} &
\colhead{Near Glitch~2} &
\colhead{Near Glitch~3} \\
}
\startdata
MJD range & 52001.684$-$52981.186 & 52773.845$-$53244.179  & 53579.360$-$53970.681 \\
TOAs & 30 & 26 & 31 \\
Epoch (MJD) & 52464.00448 & 52997.0492 & 53823.9694 \\
$\nu$ (Hz) & 0.084903950(2) & 0.084889815(3) & 0.084868657(4) \\
$\dot{\nu}$ (10$^{-13}$ Hz s$^{-1}$) & $-$2.8980(10) & $-$3.162(3) & $-$2.872(4) \\
Glitch Epoch (MJD) & 52464.00448 & 52997.0492 & 53823.9694 \\
$\Delta\nu$ (Hz) & 4.78(7)$\times$10$^{-7}$ & 2.08(4)$\times$10$^{-7}$ & 1.18(7)$\times$10$^{-7}$ \\
$\Delta\dot{\nu}$ (Hz s$^{-1}$) & $-$2.78(2)$\times$10$^{-14}$ & 4(3)$\times$10$^{-16}$ & 2(1)$\times$10$^{-15}$ \\
$\Delta \nu _{d}$ (Hz) & 8.1(6)$\times$10$^{-7}$ & $-$ & $-$ \\
$t_d$ (days) & 43(3) & $-$ & $-$ \\
RMS residual (phase) & 0.022 & 0.015 & 0.022 \\
\enddata
\tablenotetext{a} {Numbers in parentheses are TEMPO-reported 1$\sigma$
uncertainties. } 
\end{deluxetable}

\end{document}